\documentclass[%
 reprint,
superscriptaddress,
groupedaddress,
unsortedaddress,
runinaddress,
 amsmath,amssymb,
 aps,twoside,twocolumn,9pt
]{revtex4-2}

\usepackage{graphicx}
\usepackage{dcolumn}
\usepackage{bm}
\usepackage{caption}
\usepackage{subcaption}
\usepackage{fontawesome}
\usepackage{wasysym}
\usepackage[left=1.5cm, right=1.5cm, top=1.785cm, bottom=2.0cm]{geometry}
\usepackage{array}
\usepackage{droidsans}
\usepackage{charter}
\usepackage[T1]{fontenc}
\usepackage[usenames,dvipsnames]{xcolor}
\usepackage{setspace}
\usepackage[compact]{titlesec}
\usepackage{hyperref}
\usepackage{bm}

\usepackage{epstopdf}
\usepackage{blindtext}
\usepackage{lipsum}  
\usepackage{amsmath}
\usepackage{amssymb}
\usepackage{bm}
\usepackage{physics}
\usepackage{comment}
\usepackage{dcolumn}
\usepackage[normalem]{ulem}
\usepackage{xcolor}
\usepackage{caption}
\usepackage{subcaption}
\usepackage{tikz}
\usepackage{balance}
\usepackage{mathptmx}

\usepackage{graphicx} 
\usepackage{lastpage}
\usepackage[format=plain,justification=justified,singlelinecheck=false,font={stretch=1.125,small,sf},labelfont=bf,labelsep=space]{caption}
\usepackage{float}
\usepackage{fnpos}
\usepackage[english]{babel}
\addto{\captionsenglish}{%
  
}

\newcommand{\rev}[1]{{\textcolor{black}{ #1}}}

\begin{document}

\preprint{APS/123-QED}

\title{Shape, confinement and inertia effects on the dynamics of a driven spheroid in a viscous fluid}

\author{
Aditya Bhowmik\textit{$^{a}$},
Kevin Stratford\textit{$^{b}$},
Oliver Henrich\textit{$^{c\ddag}$},
Sumesh P. Thampi\textit{$^{a}$$^{\ast}$}
}
\affiliation{$^{a}$~Department of Chemical Engineering, Indian Institute of Technology Madras, Chennai 600036, India,\\
$^{b}$~EPCC, The University of Edinburgh, Edinburgh EH8 9BT, United Kingdom,\\
$^{c}$~Department of Physics, University of Strathclyde, Glasgow G4 0NG, United Kingdom.}

\thanks{Tel: 044 2257 4179; E-mail: sumesh@iitm.ac.in,}, \\
\thanks{\textit{$^{\ast}$~Corresponding author.}}\\
\thanks{\textit{$^{\ddag}$~Co-corresponding author.}}

\begin{abstract}
The dynamics of anisotropic particles in viscous flows underpin a wide range of processes in soft matter, microfluidics, and targeted drug delivery. Here, we investigate the motion of externally driven prolate and oblate spheroids suspended in a Newtonian fluid and confined within a square microchannel. Using lattice Boltzmann simulations, complemented by far-field hydrodynamic theory based on superposition of wall interactions, we systematically quantify how particle aspect ratio, strength of confinement, and fluid inertia influence the dynamics of a spheroid. For unconfined spheroids, we show that the translational velocity is maximised not for a sphere but for a prolate (end-on) or oblate (broadside-on) spheroid of specific aspect ratio. Under confinement, the optimal aspect ratio shifts toward oblate shapes due to the dominant contribution of wall-induced frictional resistance. Off-centre positioning introduces strong translation–rotation coupling, giving rise to two families of oscillatory trajectories — glancing and reversing — whose existence and structure are captured as closed orbits in the phase space. Weak fluid inertia breaks these closed loops: glancing trajectories spiral outward and merge with reversing trajectories, and new stable fixed points emerge. Together, these results reveal how modest deviations from sphericity or creeping-flow conditions profoundly alter the dynamics of driven particles in confined geometries. The predictions offer guidelines for optimising particle shape in microfluidic transport and highlight the rich nonlinear behaviour accessible in confined suspensions of nonspherical colloids.
\end{abstract}

\maketitle

\section{Introduction}

\label{sec:Intro}
Dynamics of anisotropic particles, whether driven or self-propelled, is a problem of fundamental importance in fluid mechanics and statistical physics, and relevant to various engineering applications. Because of its close connection to fluctuation dissipation theorem \cite{han2006brownian}, tracking colloidal particles forms the foundation of microrheological measurements in soft and biological matter. Recent investigations have revealed the hydrodynamic consequences emerging from heterogeneous environments such as solid walls and interfaces on the Brownian dynamics of colloids. However, these studies have predominantly focused on spherical particles \cite{fares2024observation, avni2021brownian, boniello2015brownian, molaei2021interfacial}. In contrast, the role of particle shape in these problems remains comparatively underexplored. 

Targeted drug delivery represents one of the most promising applications of driven colloids \cite{zhou2021magnetically,baulin2025intelligent, gompper20252025,patino2025swarming,ju2025technology}, yet most work to date has concentrated on spherical particles, largely due to the relative simplicity of their fabrication and their suitability as model systems \cite{he2024propulsion, gompper20252025}. Advances in synthesis methods, however, now enable the fabrication of nonspherical colloids, thereby opening new avenues of research \cite{hueckel2021total,chen2015non,lotito2022manipulating}. Recent studies have demonstrated therapeutic advantages of nonspherical particles in cancer treatment \cite{zhu2019non,ben2023shape,truong2015importance}, but their broader potential as versatile drug carriers remains insufficiently investigated. Since such particles must move in biological fluids and complex, heterogeneous environments within the body \cite{esporrin2025smart,chen2025micro,caraglio2024learning,zhang2025design}, it is essential to determine whether, and in what ways, particle shape plays a role.

Analysis of translational dynamics of a driven spheroid immersed in an unbounded, Newtonian, viscous fluid has a long history with exact solutions available in the low Reynolds number limit (Stokes flow) \cite{oberbeck1876ueber, jeffery1922motion, happel2012low}. The Reynolds number describes the ratio of inertial to viscous forces in the fluid, ${\mathcal Re} = U a/\nu$ where $U$ and $a$ are the characteristic velocity and size (the semi-major axis) of the spheroid respectively, and $\nu$ is the kinematic viscosity of the surrounding fluid.
Limiting cases of the spheroid geometry - slender bodies, spheres and discs - are most often studied, and are amenable to analytical calculations \cite{batchelor1967introduction}. Despite the long history, researchers continue to find the dynamics of spheroidal particles fascinating \cite{nissanka2023dynamics, gong2024active,sharma2025sedimentation}.

In an otherwise quiescent, infinite fluid at low Reynolds number, the dynamics of spheroids moving under an external force has two interesting characteristics \cite{guazzelli2011physical}: (i) they do not undergo rotation due to reversibility constraints, and (ii) as they translate in the direction of the applied force, they also migrate transversely (perpendicular to the applied force). However, these characteristics are not preserved in the vicinity of a solid boundary, where hydrodynamic interactions with the wall generate additional torques and transverse forces \cite{russel1977rods}. The result of the consequent coupled translational and rotational motion is oscillatory dynamics of the spheroid. Depending on the initial conditions, spheroids may undergo `glancing', `reversing' or 'periodic tumbling' dynamics \cite{mitchell2015sedimentation}. Near a slanting wall additional modes such as sliding or wobbling motion may also be observed \cite{mitchell2015sedimentation}.

Similarly, normal stresses arising from fluid inertia can exert torque on spheroids. For an unconfined spheroid, inertial torque tends to reorient its long axis perpendicular to the direction of the applied force \cite{khayat1989inertia,huang1998direct,pan2002direct,feng1994direct,dabade2015effects}. Stronger inertial effects can induce wake instabilities and unsteady modes near the translating spheroid. In the case of a confined spheroid, both inertial forces and hydrodynamic interactions with the confining boundaries act simultaneously. Such conditions commonly arise in driven colloids and particles used for targeted drug delivery, where both the degree of confinement and the Reynolds number can vary significantly. In the present work, we investigate the effect of confinement and weak inertia ($\sim\mathcal{O}(1)$) — separately and in combination — on the translational dynamics of spheroids.

Based on the analysis near a single wall, \citet{russel1977rods} suggested that spheroids would exhibit oscillatory trajectories between two parallel walls. The reversibility constraints of Stokes flow imply that such glancing-reversing trajectories \cite{mitchell2015sedimentation}  must be dependent on the initial conditions. For ${\mathcal Re} > 0$, investigations of spheroids translating in narrow tubes \cite{huang2014sedimentation,yang2015sedimentation,xia2009flow} have revealed several distinct modes of motion such as translation at constant orientation, planar oscillations and spiralling trajectories. 
These modes are found to be independent of the initial conditions. This suggests the existence of a Reynolds-number-dependent transition in the translational modes of spheroids confined within a channel. Limited numerical studies in cylindrical channels have shown Reynolds-number-dependent radial positions in the cylinder at which spheroids translate without reorienting \cite{swaminathan2006sedimentation}, although it remains unclear whether these stable solutions arise specifically from the symmetry of the cylindrical geometry.

In confinement and other complex geometries, numerical techniques are required to analyse the dynamics of spheroids. Several techniques have been developed, which include the constrained-force technique \cite{fonseca2005simulation}, arbitrary Lagrangian-Eulerian based
finite-element method \cite{swaminathan2006sedimentation,huang1998direct}, direct numerical simulation (DNS) with Lagrange multiplier based fictitious domain methodology \cite{pan2002direct, pan2005simulating}, Stokesian dynamics \cite{claeys1993suspensions}, boundary element method \cite{mitchell2015sedimentation}, immersed boundary method with DNS \cite{ardekani2016numerical}, and the lattice Boltzmann method (LBM) \cite{huang2014sedimentation, yang2015sedimentation,xia2009flow,thampi2024simulating}. 

In this work, we employ a recently developed lattice Boltzmann framework \cite{thampi2024simulating} to study a model system: the dynamics of prolate and oblate spheroids immersed in a viscous fluid and confined within a channel of square cross section. The particle is driven by an external magnetic \cite{zhang2025design} or gravitational field, which applies a force on the particle. We systematically investigate the roles of confinement and inertia on the translational dynamics of the spheroid. \rev{We consider only driven particles (by an external force) in this work which are experimentally easy to realise, say, compared to  active particles for which the leading force description is a force dipole.}  We find that for a given volume, the maximum translational velocity of an unconfined spheroid does not occur for a sphere but rather for a prolate or an oblate shape depending on its orientation relative to the applied force. In confinement, the optimal aspect ratio shifts toward oblate shapes, highlighting a promising direction for future experimental studies in drug delivery applications. Furthermore, analysis of trajectories reveals that spheroids display glancing–reversing motions depending on the initial conditions, aspect ratio, and degree of confinement. At finite Reynolds numbers, inertial nonlinearities alter this behaviour, and by systematically increasing the Reynolds number from the Stokes regime to $\mathcal{O}(1)$, we uncover the emergence of rich nonlinear dynamics in the motion of confined spheroids.

The paper is organised as follows. Section~\ref{sec:theory} introduces the model system and briefly reviews the existing literature of the analytical framework for translating spheroids in unconfined systems.~We will discuss an approximate calculation based on the method of images in section~\ref{confinement_theory}. This will be followed in section~\ref{sec:simulation} by the description of the numerical method based on lattice Boltzmann method used in this work to analyse the dynamics of confined spheroids. In section~\ref{sec:results} we will discuss the results in the following order: optimal aspect ratio for the unconfined and confined spheroids, effect of confinement on the glancing-reversing trajectories, and their phase-space description. We will end by highlighting the nonlinear features on the translational dynamics of the spheroid that emerge due to fluid inertia  and then conclude with a discussion on the scope of the work in section~\ref{sec:conclusions}.

\section{Theory}
\label{sec:theory}
\begin{figure*}
     \centering
     \begin{subfigure}[b]{0.45\textwidth}
         \centering
         \includegraphics[width=\linewidth]{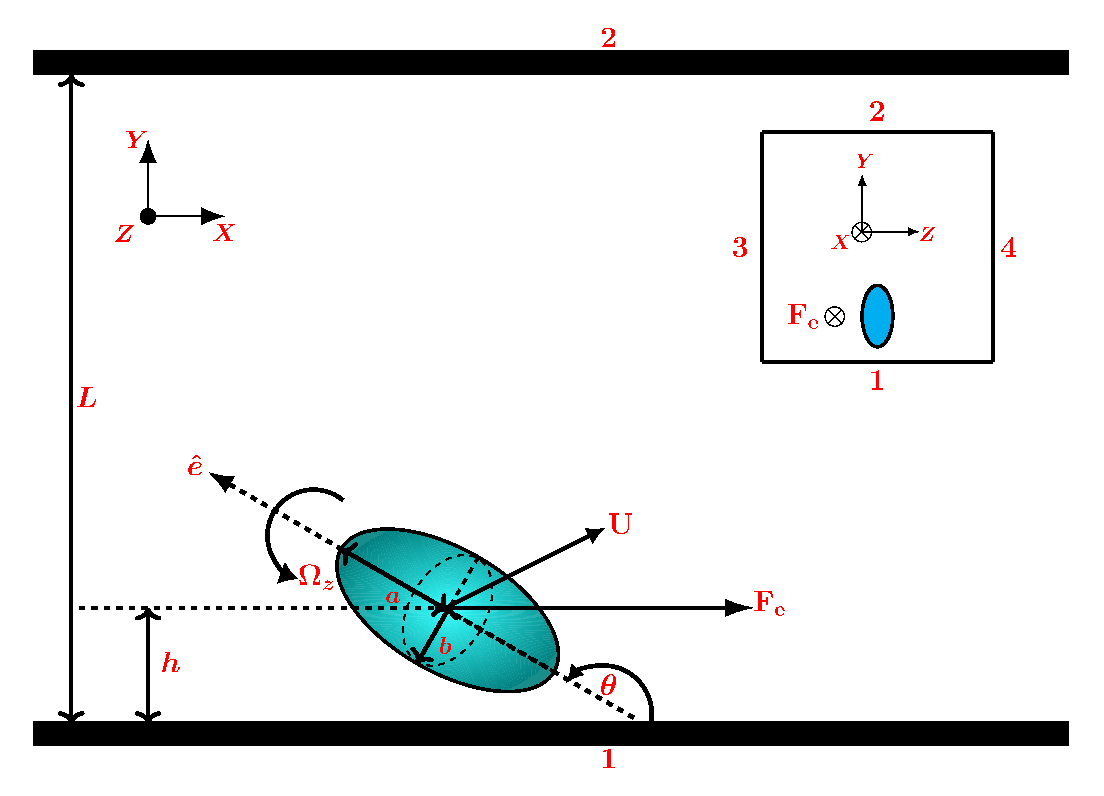}
         \caption{}
        \label{fig:prlschm}
    \end{subfigure}
    \hfill
     \centering
     \begin{subfigure}[b]{0.49\textwidth}
         \centering
         \includegraphics[width=\linewidth]{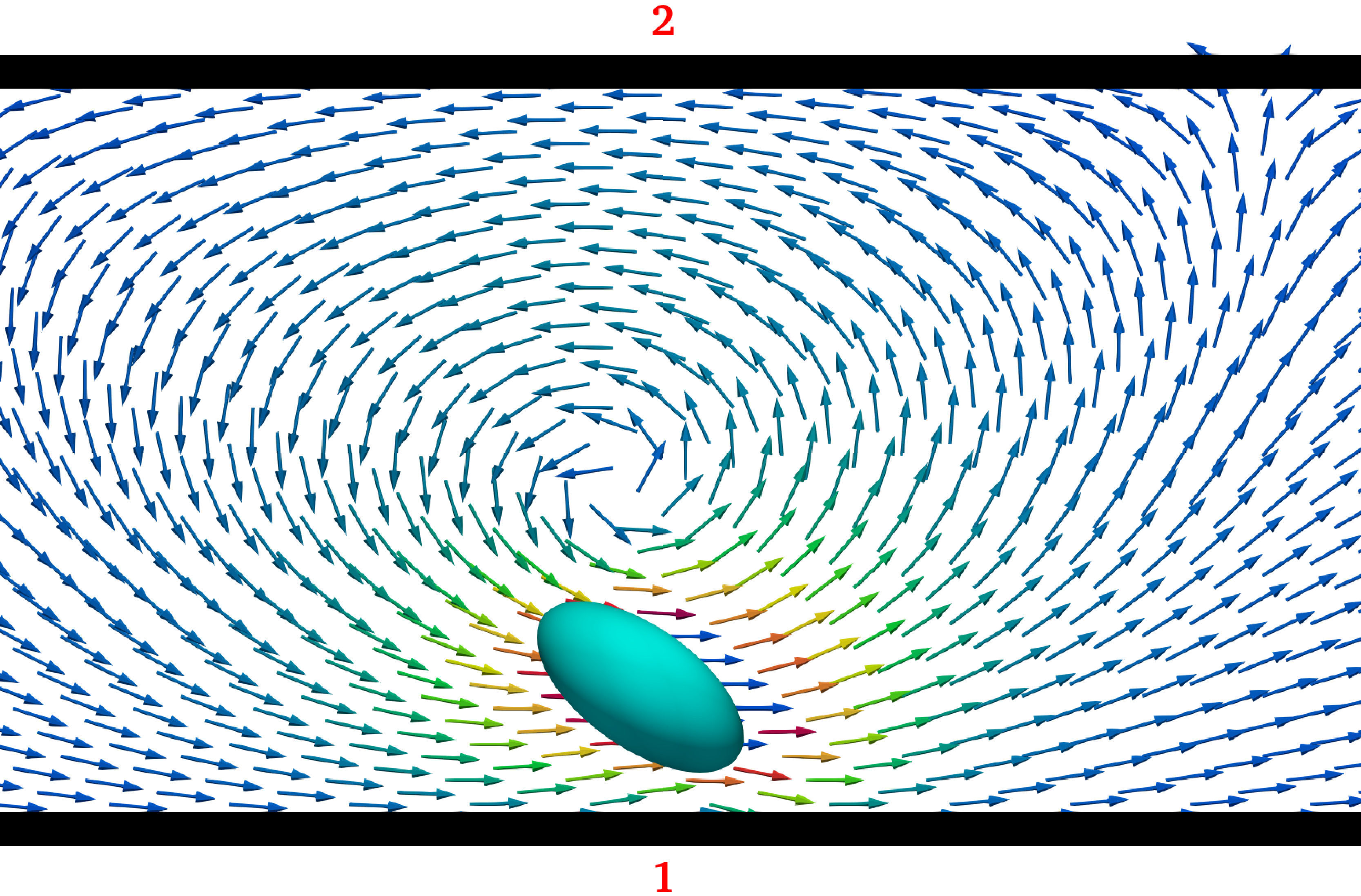}
         \caption{}
        \label{fig:flowfield}
    \end{subfigure}
    \hfill
     
     \caption{(a) Schematic representation of a translating spheroid with semi-major axis $a$ and semi-minor axes $b = c$, located at $(X, Y = h, Z = L/2)$, oriented at $\theta$, experiencing an external force $\bm{F}_e$, confined in a channel of square cross section of side length $L$. The side view of the square channel ($Y-Z$ cross section) is shown in the inset. The walls are numbered as 1, 2, 3 \& 4. (b) Fluid flow generated by the translating spheroid in the square channel, in an otherwise quiescent Newtonian fluid. The arrows indicate the velocity field, and are coloured according to the magnitude of velocity with red being the highest velocity.}
\end{figure*}

Consider a spheroid translating under the action of an external force $\bm{F}_e$ in an otherwise quiescent fluid as shown in Fig.~\ref{fig:prlschm}. The spheroid and the fluid are confined in a channel with a square cross section of side length $L$. The instantaneous orientation of the major axis of the ellipsoid is indicated by a unit vector $\hat{\bm{e}}$. Let the spheroid be of semi-major axis $a$ and semi-minor axes $b = c$. For prolate spheroids, $a > b = c$, and its eccentricity is defined as $\epsilon = \sqrt{1 - b^2/a^2}$; $0 \le \epsilon < 1$. Similarly, for oblate spheroids, $a < b = c$, and the eccentricity is  $\epsilon = \sqrt{1- a^2/b^2}$; $0 \le \epsilon < 1$. 

The fluid motion is described by the incompressible Navier-Stokes equations:
\begin{align}
 \nabla\cdot\bm{u} &= 0\\
 \rho\left(\partial_t\bm{u}+\bm{u}\cdot\nabla\bm{u}\right)&=-\nabla P+\eta\nabla^2\bm{u}
\end{align}
Here, $P$ is the pressure field, $\bm{u}$ is the fluid velocity field, $\rho$ is the density and $\eta = \nu\rho$ is the dynamic viscosity and $\nu$ is the kinematic viscosity of the fluid.

\rev{The force $\bm{F}_e$ may be any body force applied externally and  acts on the particle to drive the particle to translational motion. For example, the external force field can be considered to arise from a magnetic field when the spheroid is a microrobot, or gravitational field when gravitational sedimentation is considered.}

\rev{We consider the translation of unconfined and confined spheroids. Here, `unconfined' refers to the translation of the spheroids in the bulk fluid where wall effects are not present. `Confined' spheroids refer to particles that are enclosed within the four walls of the square channel, as shown in Fig.~\ref{fig:prlschm}, and thus the hydrodynamic interactions with the walls play a dominant role in their dynamics. The strength of the confinement is described by defining a confinement ratio, CR = $2b/L$ where $L$ is the height (equal to the depth) of the channel.}

\subsection{Translation of the unconfined spheroid}
\label{sec:unconfinedtheory}
We briefly discuss the translational dynamics of an unconfined spheroid first. At zero Reynolds number, the steady state velocity of a single, externally driven spheroid can be exactly determined~\citep{oberbeck1876ueber, chwang1976hydromechanics,happel2012low,leal2007advanced} as
\begin{align}
   \bm{F}_e &= 6 \pi \eta a (C_{\parallel} U_{\parallel}  \hat{\bm{e}} + C_{\perp} U_{\perp}  \hat{\bm{e}}_{\perp})\label{eq:settlingvel}
\end{align}
where $\hat{\bm{e}}$ and $\hat{\bm{e}}_{\perp}$ are the unit vectors along and perpendicular to the long axis of the spheroid, respectively, $U_{\parallel}$ and $U_{\perp}$ are the translational velocities along $\hat{\bm{e}}$ and $\hat{\bm{e}}_{\perp}$, respectively, \textit{i.e.,} $\mathbf{U} = U_{\parallel}\hat{\bm{e}} + U_{\perp}\hat{\bm{e}}_{\perp}$, and $C_{\parallel}$ and $C_{\perp}$ are the corresponding drag coefficients given by \citep{jiang2024settling},
\begingroup
\allowdisplaybreaks
\begin{align}
 C_{\parallel} &= \frac{8}{3} \epsilon^3 \left[ -2\epsilon + (1+\epsilon^2) \log\frac{1+\epsilon}{1-\epsilon} \right]^{-1}\label{prlcpar}\\
 C_{\perp} &= \frac{16}{3} \epsilon^3 \left[ 2\epsilon + (3\epsilon^2-1) \log\frac{1+\epsilon}{1-\epsilon} \right]^{-1}\label{prlcper}
\end{align}
\endgroup
for the prolate spheroids and 
\begingroup
\allowdisplaybreaks
\begin{align}
    C_{\parallel} =\frac{4}{3}\epsilon^3  \left[(  2\epsilon^ {2}  -1)  \cot ^ {-1}  \left(  \frac {\sqrt {1-\epsilon^ {2}}}{\epsilon} \right)+\epsilon  \sqrt {1-\epsilon^ {2}}\right]^{-1}
    \label{oblcpar}\\
    C_{\perp} =   \frac{8}{3}\epsilon^3 \left[(  2\epsilon^ {2}  +1)  \cot ^ {-1}  \left(  \frac {\sqrt {1-\epsilon^ {2}}}{\epsilon} \right)-\epsilon  \sqrt {1-\epsilon^ {2}} \right]^{-1}
    \label{oblcper}
\end{align}
\endgroup
for the oblate spheroids. Under the slender-body approximation, \textit{i.e.,} for $b/a \to 0$, Eqs.~\ref{prlcpar} - \ref{prlcper} reduce to ${C_{\parallel}} = \frac{2}{3}\frac{1}{ \ln\left(\frac{2a}{b}\right) - \frac{1}{2}}$ and ${C_{\perp}} = \frac{4}{3}\frac{1}{ \ln\left(\frac{2a}{b}\right) + \frac{1}{2}}$ respectively. Similarly, in the limit of a disc, \textit{i.e.,} $a/b\to 0$ Eqs.~\ref{oblcpar} - \ref{oblcper} reduce to $C_{\parallel} = \frac{8}{3\pi} $ and $C_{\perp} = \frac{16}{9\pi}$, respectively \citep{happel2012low}.

When the external force $\bm{F}_e$ is along the major or minor axis of the spheroid (along $\hat{\bm{e}}$ or $ \hat{\bm{e}}_{\perp}$), Eq.~\ref{eq:settlingvel} reduces to a single scalar equation relating the applied force and the corresponding translational velocity. For any other orientations of the spheroid,  Eq.~\ref{eq:settlingvel} shows that the anisotropy in the drag coefficients, \textit{i.e.,} $C_\parallel \neq C_\perp$, results in transverse forces and the migration of the ellipsoid in a direction perpendicular to the direction of its translation \cite{guazzelli2011physical}.

\subsection{Translation of the confined spheroid}
\label{confinement_theory}
Confining a force-driven spheroid in a channel can induce complex fluid flows and interesting dynamics. An example of the velocity field generated by a translating spheroid (produced using lattice Boltzmann simulations as mentioned in section~\ref{sec:simulation}) is shown in Fig.~\ref{fig:flowfield}. In other words, hydrodynamic interactions of the spheroid with confining walls alter its translational velocity and induce an angular velocity, thus leading to coupled translational-rotational dynamics of the spheroid. 

In this work, we only consider the case when $\bm{F}_e$ is along the direction of the channel length $\hat{\mathbf{x}}$ \textit{i.e.,} the externally applied force on the spheroid $\bm{F}_e$ is parallel to the confining walls.
For simplicity, we place the particle at the midplane of the channel in the $z$ direction, \textit{i.e.,} at $z = L/2$ and with long axis lying entirely in this midplane, \textit{i.e.,} the orientation vector $\hat{\bm{e}}$ in the $x-y$ plane. Owing to the symmetry of this initial condition, and despite the coupled translational-rotational dynamics, the spheroid stays and moves only in this midplane. The complete solution of this problem is obtained numerically as described in section~\ref{sec:simulation}. However, an approximate solution of the dynamics of the spheroid can be developed as follows.


Using the method of images expressions can be derived for the translational and rotational velocities of prolate and oblate spheroids placed near a wall and driven by an external force \cite{mitchell2015sedimentation}. These are far-field approximations at zero Reynolds number and near a single wall, series expressions in terms of inverse power of the distance from the wall and are valid for arbitrary orientations. Assuming a superposition principle \cite{n2024confinement}, namely that the hydrodynamic interactions of the spheroid with each wall are additive, but neglecting the corrections due to the simultaneous presence of multiple walls, and multiple reflections between the walls, we obtain
\begingroup
\allowdisplaybreaks
\begin{align}
    \bm{U} &= \bm{U}_{bulk} + \sum_{i=1}^{4} \bm{U}^{i, wall}\label{eq:Uapprox}\\
    \boldsymbol{\varOmega} &= \sum_{i=1}^{4} \boldsymbol{\varOmega}^{i,wall} \label{eq:Oapprox}
\end{align}
\endgroup
where $\bm{U}_{bulk} = U_{\parallel}\hat{\bm{e}} + U_{\perp}\hat{\bm{e}}_{\perp}$ given by Eq.~\ref{eq:settlingvel}. There are four confining walls for the channel as shown in Fig.~\ref{fig:prlschm}, and the translational and rotational velocity of the spheroid due to the hydrodynamic interaction with each wall is indicated by $\bm{U}^{i,wall}$ and $\boldsymbol{\varOmega}^{i,wall}$. They are calculated as $\bm{U}^{i,wall} = F_e/(6\pi\mu R)\tilde{\bm{U}}^{i,wall}$, $\boldsymbol{\varOmega}^{i,wall}=F_e/(6\pi\mu R^2)\tilde{\boldsymbol{\varOmega}}^{i,wall}$, where $R=a$ for prolate and $R=b$ for oblate spheroids, $F_e = |\bm{F}_e|$, and:
\begingroup
\allowdisplaybreaks
\begin{align}
    \tilde{U}_x^{1,wall} &= -\frac{9}{16h}+\frac{1}{128h^3}\left[2\epsilon^2(\cos{2\theta}\pm 1)+18\epsilon^2\cos^2{\theta}\right.\nonumber\\&-\left.(17\pm7)\epsilon^2+16\right]\label{eq:Uxfar}\\
    \tilde{U}_y^{1,wall} &= -\frac{\epsilon^2\sin{2\theta}}{32h^3}\label{eq:Uyfar}\\
    \tilde{\varOmega}_z^{1,wall} &= -\frac{9\epsilon^2(2-4\cos^2{\theta})}{64(2-\epsilon^2)h^2} + 6\left[-9\epsilon^4\cos^4{\theta} -\epsilon^2\cos^2{\theta}(12\right. -(15\nonumber\\&\pm2)\epsilon^2)-\left.(7\pm1)\epsilon^4+10\epsilon^2-4\right]/[128(2-\epsilon^2)h^4]\label{eq:Ozfar}.
\end{align}
\endgroup
The corresponding expressions for the wall on the right ($i=2$) can be obtained by appropriate substitutions:
\begingroup
\allowdisplaybreaks
\begin{align}
\tilde{U}_x^{2,wall} &= \tilde{U}_x^{1,wall}(h\rightarrow L-h,\theta\rightarrow-\theta)\\
\tilde{U}_y^{2,wall} &= -\tilde{U}_y^{1,wall}(h\rightarrow L-h,\theta\rightarrow-\theta)\\
\tilde{\varOmega}_z^{2,wall} &= -\tilde{\varOmega}_z^{1,wall}(h\rightarrow L-h,\theta\rightarrow-\theta).
\end{align}
\endgroup
The hydrodynamic interaction with the remaining pair of walls ($i=3, 4$) are given by,
\begingroup
\allowdisplaybreaks
\begin{align}
        \tilde{U}_x^{3,wall} &= \tilde{U}_x^{4,wall} = -\frac{9}{8L}\nonumber\\ &+ \frac{1}{16L^3}\left[ 2\epsilon^2(1\pm1)\cos^2{\theta}-(\pm7-1)\epsilon^2 + 16 \right]\\
    \tilde{U}_y^{3,wall} &= \tilde{U}_y^{4,wall} = 0\\
    \tilde{\varOmega}_z^{3,wall} &= \tilde{\varOmega}_z^{4,wall} = 0 .\label{eq:lastinapprox}
\end{align}
\endgroup

\rev{Analysing Eqs.~\ref{eq:Uxfar}-\ref{eq:Ozfar} we can see that the far-field prediction is $\mathcal{O}((a/h)^3)$ for translational and $\mathcal{O}((a/h)^4)$ for angular velocity.} Eqs.~\ref{eq:Uapprox} - \ref{eq:lastinapprox} can be used to determine approximately the coupled translational-rotational dynamics of the spheroid. The complete dynamics can be captured using numerical simulations as described in the next section.

\section{Simulation Method}
\label{sec:simulation}
We use the lattice Boltzmann method to determine the dynamics of the fluid around the translating spheroid confined in the channel. Through the boundary conditions the motion of the spheroid and the fluid are completely coupled. The algorithm is implemented in our parallel lattice Boltzmann code \textit{Ludwig} \citep{desplat2001ludwig, Ludwig}. We briefly discuss the method below, and refer the reader to \citet{thampi2024simulating} for a detailed discussion of the implementation.

\subsection{Lattice Boltzmann method}
The lattice Boltzmann method (LBM) defines a discrete velocity distribution function $f_i(\bm{x},t)$ at each grid point $\bm{x}$ and time $t$ that travels along the $i^{th}$ discrete direction with velocity $\bm{c}_i$. The hydrodynamic observables, namely the density $\rho(\bm{x},t)$ and momentum  $\rho(\bm{x},t)\bm{u}(\bm{x},t)$ are obtained as the zeroth and first moment of the discrete distribution function:
\begin{align}
    \sum_i f_i (\bm{x},t) &= \rho(\bm{x},t) \\
     \sum_i f_i (\bm{x},t) \bm{c}_{i} &= \rho(\bm{x},t) \bm{u}(\bm{x},t).
\end{align}

In this work, the $D3Q19$ model is used, which discretises configuration and momentum space with a 3D cubic lattice with 19 discrete lattice vectors $\bm{c}_i$ \citep{kruger2017lattice}. A lattice spacing of $\Delta x = 1$ and time step $\Delta t = 1$ are used for the spatial and temporal discretisation, respectively. 

The discrete distribution functions $f_i$ undergo successive collision and propagation operations:
\begin{align}
f_i(\bm{x} + \bm{c}_i\Delta t, t + \Delta t) - f_i(\bm{x}, t) = \Delta t \, {\cal C}_i(t).
\end{align}
where in its simplest Bhatnagar–Gross–Krook form \cite{bhatnagar_collision_1954} the collision operator contains a single relaxation time $\tau$ and is given as
\begin{equation}
    {\cal C}_i(t) = [f_i(\bm{x}, t) - f_i^{eq}(\bm{x}, t)]/\tau
\end{equation}
The kinematic viscosity $\nu$ of the fluid is related to the relaxation time via $\nu = (\tau - \frac{1}{2})/3$. The equilibrium distribution:
\begin{align}
    f_i^{eq} = \rho w^{\bm{c}_i} \left( 1 + \frac{\bm{u}\cdot\bm{c}_i}{c_s^2} + \frac{(\bm{u}\cdot\bm{c}_i)^2}{2c_s^4} - \frac{\bm{u}\cdot\bm{u}}{2 c_s^2} \right)
\end{align}
is constrained by the governing equations to be recovered, namely the Navier-Stokes equations. Here, $c_s$ is the speed of sound and $w^{\bm{c}_i}$ are the weights associated with the velocity set $\bm{c}_i$ of the $D3Q19$ model. We use a generalised, multiple relaxation time (MRT) collision operator in our simulations \cite{adhikari_fluctuating_2005}.

\subsection{Boundary conditions}
\label{sec:boundary conditions}
The surface of the spheroid is described as
\begin{align}
(\bm{x}-\bm{x}_c)^T\bm{A}(\bm{x}-\bm{x}_c) = 1,
\label{eq:ellipsoidsurface}
\end{align}
where $\bm{x}_c$ is the centre of mass, $\bm{x}$ is any point on the surface of the spheroid and $\bm{A}$ is a $3 \times 3$ matrix whose eigenvalues are related to the inverse of the semi-axes of the spheroid \citep{strang2022introduction}. Bounce-back boundary conditions are applied on the boundary nodes $\bm{x}_b = \bm{x} + \frac{1}{2}\bm{c}_b\Delta t$, which is an approximate surface of the spheroid  \citep{thampi2024simulating}.

The bounce back scheme results in momentum exchange between the fluid and the solid nodes. The total force $\bm{F}$ and torque $\bm{T}$ on the spheroid due to momentum exchange can be calculated by adding up the contributions from all boundary nodes as $\bm{F}=\sum_b \bm{F}_b$ and  $\bm{T} = \sum_b (\bm{x}_b - \bm{x}_c) \times \bm{F}_b$, where \citep{nguyen2002lubrication}
\begin{align}
    \bm{F}_b(\bm{x}_b,t+\frac{1}{2}\Delta t) = \frac{\Delta x^3}{\Delta t} \left[ 2f_b^*(\bm{x},t) - \frac{2 w^{\bm{c}_b} \rho_0 \bm{u}_b\cdot\bm{c}_b}{c_s^2}\right] \bm{c}_b.
    \label{eq:fb}
\end{align}
Here, $f_b^*(\bm{x},t)$ is the post-collision distribution function, $\rho_0$ is the mean density of the fluid, and $\bm{u}_b$ is the local velocity of the boundary node calculated as 
\begin{align}
\bm{u}_b = \bm{U} + \bm{\varOmega} \times (\bm{x}_b - \bm{x}_c).
 \label{eq:ub}
\end{align} 
where, $\bm{U}$ and $\bm{\varOmega}$ are the translational and angular velocities of the spheroid.

\subsection{Dynamics of the spheroid}
\rev{An implicit numerical scheme is used to update the translational and angular velocities of the spheroid from the force and torque determined in the previous section, accounting for particle inertia.} In this scheme \citep{nguyen2002lubrication}, the total force and torque on the particle are written as,
\begingroup
\allowdisplaybreaks
\begin{align}
\bm{F} &= \bm{F}_0 - \boldsymbol{\zeta}^{FU} \cdot \bm{U} - \boldsymbol{\zeta}^{F\varOmega} \cdot \boldsymbol{\varOmega}\label{eq:Fsplit}\\
\bm{T} &= \bm{T}_0 - \boldsymbol{\zeta}^{TU} \cdot \bm{U} - \boldsymbol{\zeta}^{T\varOmega} \cdot \boldsymbol{\varOmega}\label{eq:Tsplit}
\end{align}
\endgroup
where $\bm{F}_0$ and $\bm{T}_0$ are `velocity-independent' forces and torques, determined from the post-collision distributions and $\boldsymbol{\zeta}^{**}$ are the drag coefficient matrices \citep{thampi2024simulating}. 

Then, the discretised conservation equations of linear and angular momentum of the translating spheroid read as
\begingroup
\allowdisplaybreaks
\begin{align}
M \frac{\bm{U}(t+\Delta t) - \bm{U}(t)}{\Delta t} &= \bm{F}_0 (t + \frac{1}{2} \Delta t) \nonumber\\&- \boldsymbol{\zeta}^{FU} \cdot \bm{U}(t + \Delta t) - \boldsymbol{\zeta}^{F\varOmega} \cdot \boldsymbol{\varOmega}(t + \Delta t) \label{eq:Fdiscrete}\\
\bm{I}(t) \cdot \frac{\boldsymbol{\varOmega}(t+\Delta t) - \boldsymbol{\varOmega}(t)}{\Delta t} &+ \frac{d\bm{I}}{dt} \cdot  \boldsymbol{\varOmega}(t+\Delta t) = \bm{T}_0 (t + \frac{1}{2} \Delta t) \nonumber\\&- \boldsymbol{\zeta}^{TU} \cdot \bm{U}(t + \Delta t) - \boldsymbol{\zeta}^{T\varOmega} \cdot \boldsymbol{\varOmega}(t + \Delta t) \label{eq:Tdiscrete}
\end{align}
\endgroup
where $M$ and $\bm{I}$ are respectively the mass and moment of inertia tensor of the spheroid. A Gaussian elimination method is used to solve Eqs.~\ref{eq:Fdiscrete} and \ref{eq:Tdiscrete} simultaneously, thus ensuring the stability of the algorithm.

The position of the particle is then updated as
\begin{align}
\bm{x}_c(t + \Delta t) = \bm{x}_c (t) + \frac{1}{2}\left( \bm{U}(t + \Delta t) + \bm{U}(t) \right).
\label{eq:posupdate}
\end{align}
The orientation of the particle is updated as \citep{zhao2013direct, zhao2013novel, van2015modelling}
\begin{align}
&\tilde{\mathbf{q}} (t) = \left[\cos{\frac{||\tilde{\boldsymbol{\varOmega}} (t)||\Delta t}{2}},\sin{\frac{||\tilde{\boldsymbol{\varOmega}} (t))||\Delta t}{2}}\frac{\tilde{\boldsymbol{\varOmega}}(t))}{||\tilde{\boldsymbol{\varOmega}}(t))||}\right]\label{eq:qnplus0}\\
&\mathbf{q}(t+\Delta t) = \tilde{\mathbf{q}}(t)\mathbf{q}(t) ,
\label{eq:qnplus1}
\end{align}
where $\mathbf{q}$ is the unit quaternion describing the orientation of the spheroid based on Euler angles   \citep{voth2017anisotropic,fan1995sublayer}, $
\tilde{\boldsymbol{\varOmega}}(t) = \frac{1}{2} \left(\boldsymbol{\varOmega}(t) + \boldsymbol{\varOmega}(t + \Delta t) \right)$ is the mean angular velocity and $||\cdot||$ indicates the norm of the vector. This procedure avoids accumulated errors from successive matrix operations and singular matrix operations, which could arise if Euler angles were used to describe the orientation of the spheroid. It also avoids any possibility of a gimbal lock and requirement of renormalisation of quaternions to account for errors from numerical integration since the magnitude of the quaternion is exactly unity. 

Finally, we note that in the implicit solution scheme in Eq.~\ref{eq:Tdiscrete} the rate of change of the moment of inertia of the spheroid is not zero in the laboratory coordinate system, which is why this calculation requires careful consideration. The time-dependent moment of inertia tensor $\bm{I}(t)$ can be determined from the quaternion $\mathbf{q}(t)$,
\citep{fan1995sublayer, zhao2013direct, van2015modelling}, 
\begin{align}
\mathbf{I}(t) = \left(\mathbf{q} \left(\mathbf{q}\,\mathbf{I}\,\mathbf{q}^{-1}\right)^T\mathbf{q}^{-1}\right)^T ,
\label{eq:Irot}
\end{align}
where $\mathbf{I}(t)=[0,\bm{I}(t)]$ is a pure quaternion moment of inertia tensor based on the inertia tensor in the principal coordinate system. In the above, $^{-1}$ and $^{T}$ represent the inverse and transpose operations respectively. Then the time derivative of the moment of inertia tensor in the second term in Eq.~\ref{eq:Tdiscrete} can be exactly determined. 

\subsection{Summary of the simulation method}
The numerical procedure described in the previous section proceeds as follows. The lattice Boltzmann streaming and collision steps advance the fluid field around the moving spheroid. The mid-grid bounce-back condition is imposed on the particle surface, enabling momentum transfer between the spheroid and the surrounding fluid. This exchange enforces the no-slip boundary condition on both the spheroid and the channel walls while simultaneously providing the hydrodynamic force and torque acting on the particle. The resulting translational and rotational dynamics of the spheroid are then obtained by solving the linear and angular momentum balance equations, with the particle orientation updated using quaternions and standard quaternion operations. More details on the implementation of the algorithm and validation of the method have been reported previously \cite{thampi2024simulating}.

The simulation parameters used in this study are summarised below. For the unconfined cases, a cubic computational domain of size $256^3$ with periodic boundary conditions applied on all boundaries is employed. The major and minor axes of the spheroid vary between $6 - 60$ and $2 - 6$, respectively. \rev{We have used particles with density ratio $\rho_p/\rho_f = 1$ in the simulations with an external body force specified using $\bm{F}_e$.} To investigate the role of geometric confinement, simulations are carried out in a domain of size $256\times N_y \times N_z$, where $N_y = N_z$ is varied from $5 - 154$ to vary the strength of confinement. \rev{When a spheroid is in close proximity to a wall, the hydrodynamic lubrication between the two can be under-represented on the discrete lattice. In this case a correction to the lubrication is made based on the surface separation at the point of closest approach. This prevents the spheroid touching the wall. Our results are independent of the specific implementation of this contact force.} Periodic boundary conditions are applied on the open sides of the domain. 
To examine the influence of inertia, the kinematic viscosity of the fluid is varied $0.007 - 0.5$. In all cases, simulations are initialised with a quiescent fluid. Each run is continued until a steady state is reached, and the corresponding results are presented below.

In the Stokes flow regime, quasi-static simulations were employed wherever feasible to improve computational efficiency. In this approach, the fluid field is evolved alongside the particle velocities, but the particle is kept fixed in space. Holding the particle stationary eliminates unsteady effects and allows the hydrodynamic calculations to converge rapidly generating the instantaneous kinematic state for that configuration. 

\section{Results and Discussion}
\label{sec:results}
We consider the dynamics of both prolate and oblate spheroids by varying the aspect ratio $b/a \rightarrow 0 $ to $b/a \rightarrow \infty$. Below, we first consider the case of unconfined spheroids and demonstrate the existence of an optimal aspect ratio that yields the maximum translation velocity. We will then present the results of confined spheroids, and analyse the shift in the optimum aspect ratio as a function of strength of confinement. Furthermore, we present results that demonstrate the effects of confinement and inertia on the trajectories.


\begin{figure}[]
         \centering
         \includegraphics[trim = 0 0 40 40, clip, width=0.945\linewidth]{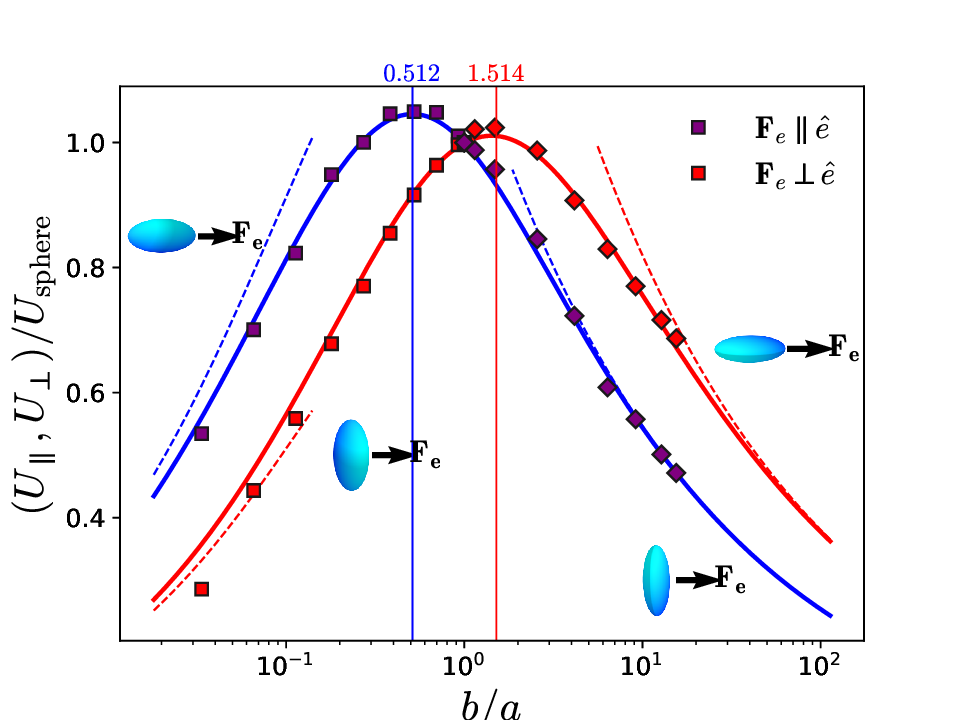}
    \caption{\rev{Steady state translational velocity of an unconfined spheroid as a function of aspect ratio for a fixed particle volume}. Aspect ratios $b/a < 1$ correspond to prolate spheroids, while $b/a> 1$ represent oblate shapes. The continuous curves show the analytical predictions (Sec.~\ref{sec:unconfinedtheory}), the symbols denote the lattice Boltzmann simulation results, and the dashed lines indicate the asymptotic expressions for rod-like and disc-like limits. \rev{In the simulations, the spheroids were initially placed at the centre of the domain, with their symmetry axis either parallel ($\parallel$) or perpendicular ($\perp$) to the applied force ($\bm{F}_e = 0.05$ lattice units). The spheroids retained this orientation through out the simulations. ${\mathcal Re}$, calculated based on the steady state translational velocity of the spheroid, $<< 1$. For the sphere ($b/a = 1$), ${\mathcal Re} \approx0.023$.} Data plotted in blue (red) correspond to end-on (broadside-on) configurations, respectively. The aspect ratio at which the translational velocity is maximum in each case is marked by a vertical line.}
         \label{fig:Ellipsoid_bulk}
\end{figure}

\begin{figure*}[t]
    \centering
    \begin{subfigure}[b]{0.32\textwidth}
        \centering
        \includegraphics[width=\textwidth]{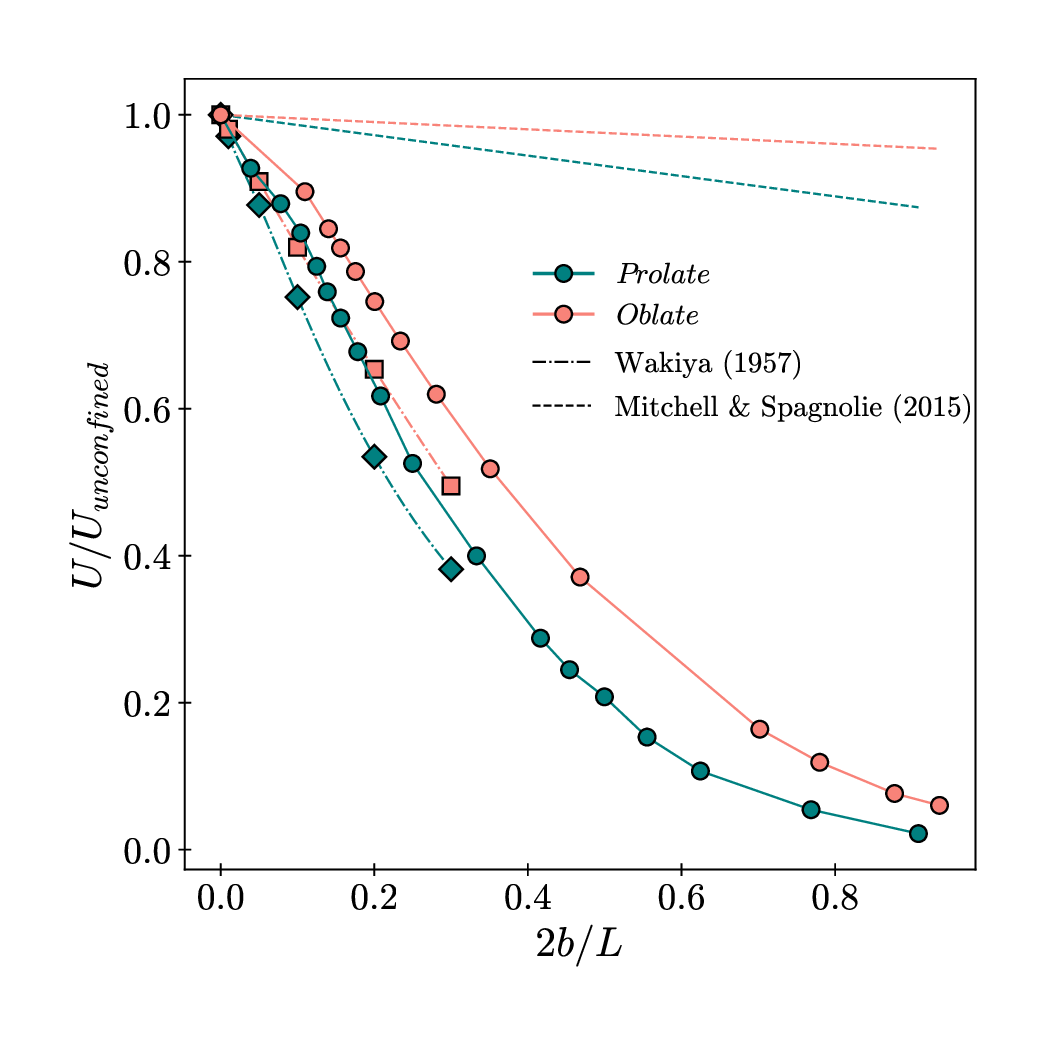}
        \caption{}
        \label{wakiya}
    \end{subfigure}
    \hfill
    \begin{subfigure}[b]{0.32\textwidth}
        \centering
        \includegraphics[width=\textwidth]{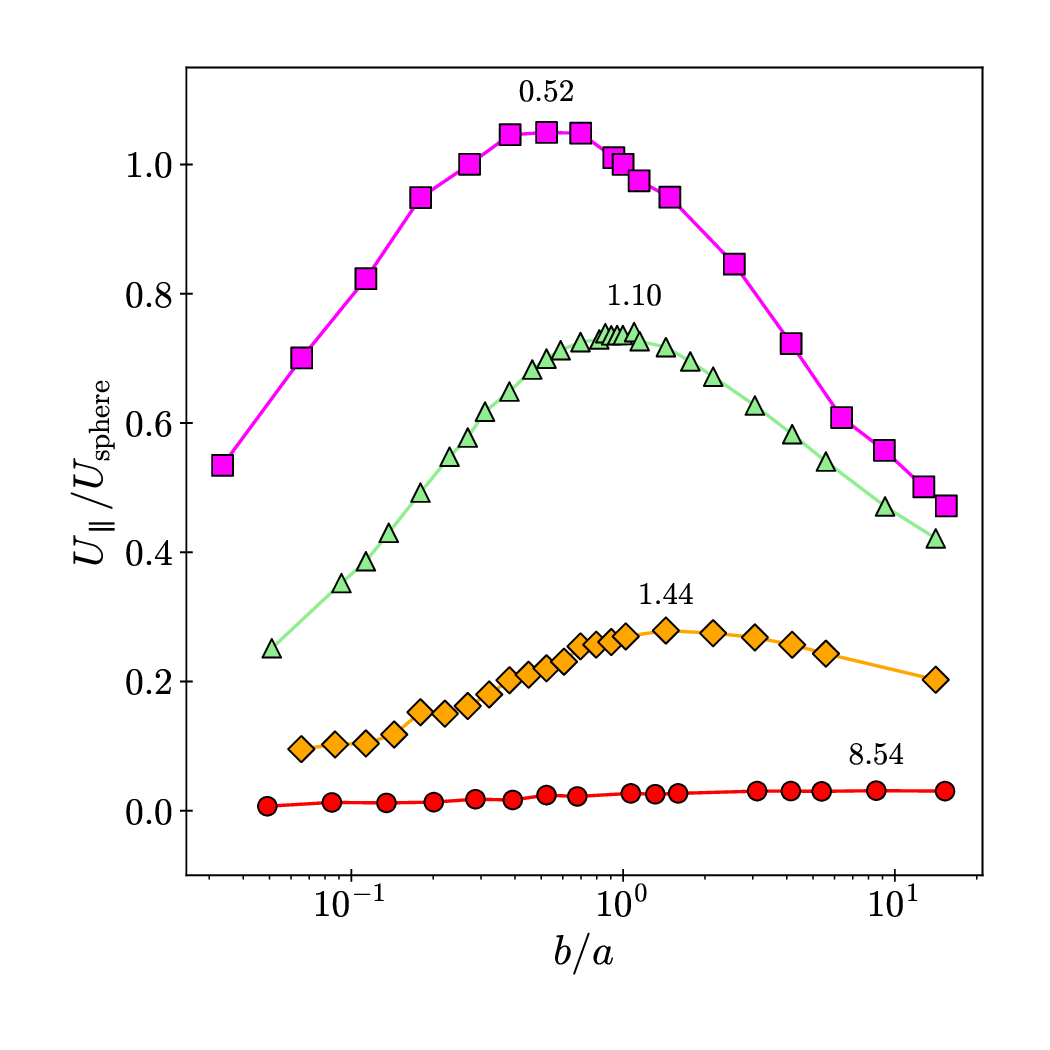}
        \caption{}
        \label{Prolate_confined}
    \end{subfigure}
    \hfill
    \begin{subfigure}[b]{0.32\textwidth}
        \centering
        \includegraphics[width=\textwidth]{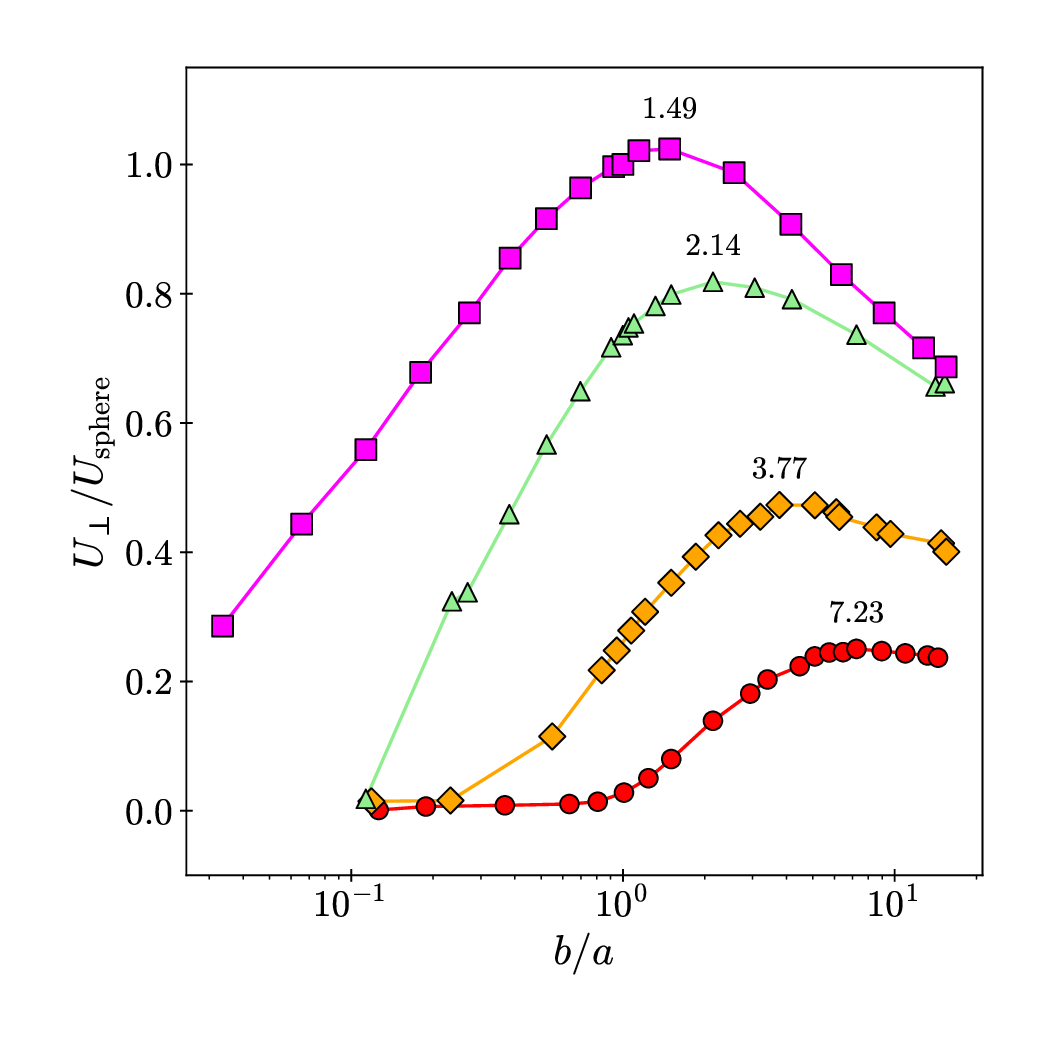}
        \caption{}
        \label{Oblate_confined}
    \end{subfigure}
    

    
    \caption{(a) Translational velocity of spheroids (prolate, $b/a~\approx0.51$ and oblate, $b/a~\approx1.57$) moving in an end-on configuration along the centreline of a square channel, shown as a function of increasing confinement ratio, CR $= 2b/L$. The symbols $\CIRCLE$ are from the LBM simulations, the dash-dotted lines are spheroids in cylindrical channels (prolate, $b/a = 0.5$ and oblate $b/a = 2$)\cite{wakiya1957viscous} and dotted lines are theoretical predictions based on far-field analysis (section~\ref{confinement_theory}). The ordinate is normalised with the translational velocity of the unconfined spheroid. Translational velocities of spheroids for (b) $\bm{F} \parallel \hat{\bm{e}}$ and (c) $\bm{F} \perp \hat{\bm{e}}$ configuration as a function of aspect ratio for fixed confinement ratios - $\blacksquare$: CR$= 0$, $\blacktriangle$: CR$= 0.2$, $\blacklozenge$: CR$= 0.5$ and $\CIRCLE$: CR$=0.9$. The ordinate is normalised with the translational velocity of an unconfined sphere and the optimum aspect ratio at each confinement ratio is indicated on top of each peak in (b) \& (c).} 
    \label{fig:settling}
\end{figure*}

\subsection{Optimum aspect ratio of an unconfined spheroid}
\label{optimus_prime}
Under the action of the external force $\bm{F}_e$, the spheroid accelerates, but the drag force generated by the fluid opposes this motion. Balancing the two opposing forces, the particle undergoes a steady translation with a constant velocity. The variation in the translational velocity of unconfined prolate and oblate spheroids in an otherwise quiescent fluid as a function of aspect ratio $b/a$ is shown in Fig.~\ref{fig:Ellipsoid_bulk}.

The aspect ratio $b/a < 1$ for prolate and $>1$ for oblate spheroids. Two different cases are considered in Fig.~\ref{fig:Ellipsoid_bulk}: (i) the end-on translation, \textit{i.e.,} $\hat{\bm{e}} \parallel \bm{F}_e$, the spheroid translates with its axis of revolution oriented $parallel$ to the direction of the external force and (ii) the broadside-on translation, \textit{i.e.,} $\hat{\bm{e}} \perp \bm{F}_e$, the spheroid translates with its axis of revolution oriented $perpendicular$ to the direction of the external force. The corresponding velocities are denoted as $U_\parallel$ and $U_\perp$ respectively. The symbols are obtained from lattice Boltzmann simulations and the continuous lines are obtained from the analytical expression, Eq.~\ref{eq:settlingvel}. The dashed lines on the extremes of both sides of the aspect ratio are analytical approximations - based on slender bodies for prolate and discs for oblate spheroids.

In typical applications driven by external fields such as magnetic or gravitational, the force density (force per unit volume) on the particle will be constant. Hence, in Fig.~\ref{fig:Ellipsoid_bulk} the aspect ratio of the spheroid is varied by keeping the volume of the spheroid $V = \frac{4}{3}\pi ab^2$ constant so that the total force acting on the translating particle is fixed. Furthermore, the translational velocities, $U_\parallel$ and $U_\perp$ are non-dimensionalised with $U_{sphere} = \frac{F_e}{6\pi\eta}(\frac{4\pi}{3V})^{1/3}$, the Stokes velocity of an unconfined sphere of same volume $V$. Thus, $U_\parallel/U_{sphere} = U_\perp/ U_{sphere} = 1$ when $b/a = 1$.  

The excellent match between the simulation data and the analytical solutions at all aspect ratios indicates the reliability of the numerical method used in this work. More importantly, Fig.~\ref{fig:Ellipsoid_bulk} shows the existence of a maximum in the two cases considered. For the case of end-on translation (blue curve) the maximum in the translational velocity occurs at an aspect ratio $b/a \approx 0.512$, that of a prolate shape. For the case of broadside-on translation (red curve) the maximum occurs at an aspect ratio of $b/a\approx 1.514$, that of an oblate shape. The curves corresponding to two cases intersect at $b/a = 1$, that of the sphere, and the maxima in two cases occur on either side of the sphere.

The maximum in the translational velocity (or minimum in the drag force) has been observed for a variety of anisotropic particles in the context of gravitational sedimentation \cite{carmichael1982estimation, happel2012low, pourali2021drag, loth2008drag, sanjeevi2022accurate}. It exists because the viscous drag force acting on the translating particle has two contributions (i) friction drag and (ii) pressure drag, and these two contributions scale differently with change in the dimensions of the particle \cite{aoi1955steady, leith1987drag, ouchene2024orientation}. The skin friction arises from the shear stress on the particle's surface due to the viscosity of the fluid, and this contribution depends on the surface area of the particle. The  pressure drag, also known as the form drag, arises as the particle pushes the fluid in front of it to move ahead. Thus, there is a pressure gradient across the translating particle, and the pressure drag contribution primarily depends on the projected area of the particle in the direction of translation \cite{leith1987drag}. The difference in the dependence of friction and pressure drag on the spheroid surface area results in the presence of maxima in Fig.~\ref{fig:Ellipsoid_bulk}, and can be rationalised as follows.

Consider a sphere of radius $r$ translating under an external force. Both the total surface area, and the projected surface area scale as $\sim r^2$. Exact calculations show that the friction drag contributes $2/3^{\textnormal{rd}}$ and the pressure drag contributes $1/3^{\textnormal{rd}}$ of the total drag force \cite{aoi1955steady, leith1987drag}.

To examine a small deviation from sphericity, let us consider a prolate spheroid with $a > r$. For end-on translation, the projected surface area is $\sim b^{2}$ while the total (or lateral) surface area is $\sim ab$. Under the constraint of fixed volume $V \sim ab^{2}$, the projected area scales as $\sim V/a$ and the total area as $\sim \sqrt{Va}$. Thus, changing the shape from a sphere to a prolate spheroid decreases the projected surface area as $\sim 1/a$ but increases the total surface area as $\sim \sqrt{a}$.
As a result, the reduction in pressure drag outweighs the increase in friction drag, leading to a net decrease in the total drag force; consequently, an end-on translating prolate spheroid attains a higher velocity than a sphere of equal volume.
 Exact computations confirm this scaling argument \cite{ouchene2015drag, ouchene2024orientation}.



This trend however does not hold for highly elongated prolate spheroids ($a >> b$), where the lateral area far exceeds the projected area, \textit{i.e.} $ab >> b^{2}$. In this regime, friction drag dominates, as seen in analytical results for needle-like bodies where pressure drag vanishes \cite{aoi1955steady} and in numerical studies of spheroids \cite{ouchene2024orientation}. Thus, for highly elongated particles, the total drag increases with $a$.

In other words, prolate spheroids that are closer to a sphere in shape ($b/a \lesssim 1$) translate faster on increasing the length of the major axis, $a$, but a similar change of increasing $a$ slows down highly elongated, rod-like spheroids ($b/a << 1$). This contrast in the dependence of drag force at large ($b/a \lesssim 1$) and small ($b/a << 1$) aspect ratios gives rise to a maximum in the translational velocity at an intermediate value of aspect ratio for the translating prolate spheroid in the end-on configuration. 

By contrast, when a sphere is deformed into an oblate spheroid translating in the end-on configuration, the increase in projected surface area  ($\sim b^2$)  exceeds the decrease in lateral area ($\sim ab$), leading to a reduction in translational velocity relative to that of a sphere. This trend persists at larger aspect ratios ($b/a >> 1$), so that the translational velocity of an oblate spheroid decreases monotonically from the spherical limit ($b/a \gtrsim 1$). Similar arguments of anisotropic scaling of the relevant surface areas hold for broadside on translation as well, and a maximum velocity appears for an oblate spheroid in this configuration.

\subsection{\textbf{Confined spheroid: centreline of the channel}}
\label{sec:centre_confined}

In this section, we investigate the influence of confinement on the translational velocity of a spheroid by placing it inside a channel with a square cross‐section. As illustrated in Fig.~\ref{fig:prlschm}, the external force is applied along the open direction of the channel (the $x-$axis). For simplicity, the spheroid is positioned on the channel centreline in an end‐on configuration, \textit{i.e.}, its axis $\hat{\bm{e}}$ is aligned with the channel axis. \rev{As in the previous section, we apply a constant body force density to the spheroid by keeping the volume of the spheroid and the applied body force $\bm{F}_e$ constant while changing the aspect ratio and the confinement ratio.} 


The effect of confinement is quantified using the confinement ratio, CR = $2b/L$ where $L$ is the height (equal to the width) of the channel. As shown in Fig.~\ref{wakiya}, the translational velocity of both prolate and oblate spheroids decreases with increasing confinement. Prolate spheroids exhibit lower translational velocities than oblate spheroids because they share a larger surface area with the channel walls and therefore experience greater hydrodynamic resistance.

Figure \ref{wakiya} also includes translational‐velocity data for spheroids (i) in a cylindrical channel and (ii) obtained from the approximate analytical method for comparison. \citet{wakiya1957viscous} analysed the axisymmetric translation of spheroids in cylindrical channels using a combined analytical–numerical approach, and reported the translational velocity accurate up to ~$\mathcal{O(\textnormal{CR})}^4$ in a tabulated form. Data extracted from his results for comparable aspect ratios — although available for only a few confinement ratios (CRs) — are shown in Fig.~\ref{wakiya} as dash-dotted lines. The slightly larger translational velocities observed in the simulations arise from geometric differences; specifically, a square channel is marginally “less confining’’ than a cylindrical one due to the presence of corners.
The results from the approximate analytical calculations, accurate up to ~$\mathcal{O(\textnormal{CR})}^3$ and based on the superposition arguments discussed in Section~\ref{confinement_theory}, are also shown in Fig.~\ref{wakiya} as dotted lines. These results systematically over-predict the translational velocity and exhibit only qualitative agreement with the simulations.



We now examine how the translational velocity of confined spheroids varies with their aspect ratio. Figure~\ref{Prolate_confined} shows the normalised translational velocity of spheroids moving along the channel centreline in the end-on configuration, with aspect ratios ranging from $b/a < 1$ (prolate) to $b/a > 1$ (oblate). As in the unconfined case, we find that for each confinement ratio there is an aspect ratio at which the translational velocity reaches a maximum. This optimum shifts to higher aspect ratios (toward the oblate side) as the degree of confinement increases.

The shift of the maximum translational velocity toward the oblate side arises from the nature of the dominant drag mechanisms. In this geometry, the primary contribution to hydrodynamic resistance comes from friction along the lateral surface of the particle, generated by the shear flow between the spheroid and the channel walls. For a given confinement ratio and in the end-on configuration, an oblate spheroid presents a smaller lateral surface area to the walls than a prolate one, leading to lower frictional drag, and therefore a higher translational velocity. However, as the aspect ratio becomes very large, the gap between the particle and the walls becomes quite small. This leads to strong shear and a substantial increase in frictional drag. Consequently, the total drag becomes very large as $b/a>>1$. This implies that the translational velocity must peak at a finite aspect ratio, and that this optimum lies on the oblate side.

We also consider the broadside-on configuration of spheroids translating along the channel centreline. The corresponding results are presented in Fig.~\ref{Oblate_confined}. As in the previous cases, an optimal aspect ratio exists; it occurs for an oblate particle and shifts to larger values as the confinement increases. In this configuration, the location of the optimum is again determined by the frictional resistance between the particle and the channel walls, but for a different geometric reason. As the aspect ratio of an oblate spheroid increases, its flatter faces move farther from the walls while its thinner edges move closer. Consequently, the hydrodynamic resistance initially decreases at first because the dominant flat surfaces experience lower frictional forces, before increasing again at very large aspect ratios due to the progressively narrower gaps near the thin edges. The curves tend to flatten at very high aspect ratios, as the contribution of these thin edges to the overall frictional drag becomes relatively constant.

Overall, these results indicate that an optimal aspect ratio exists for maximising translational velocity in both confined and unconfined settings. For most configurations, this optimum lies on the oblate side, where oblate spheroids achieve higher translational velocities than spheres and prolate counterparts.

\begin{figure*}
     \centering
     \begin{subfigure}[b]{0.33\textwidth}
         \centering
         \includegraphics[width=\linewidth]{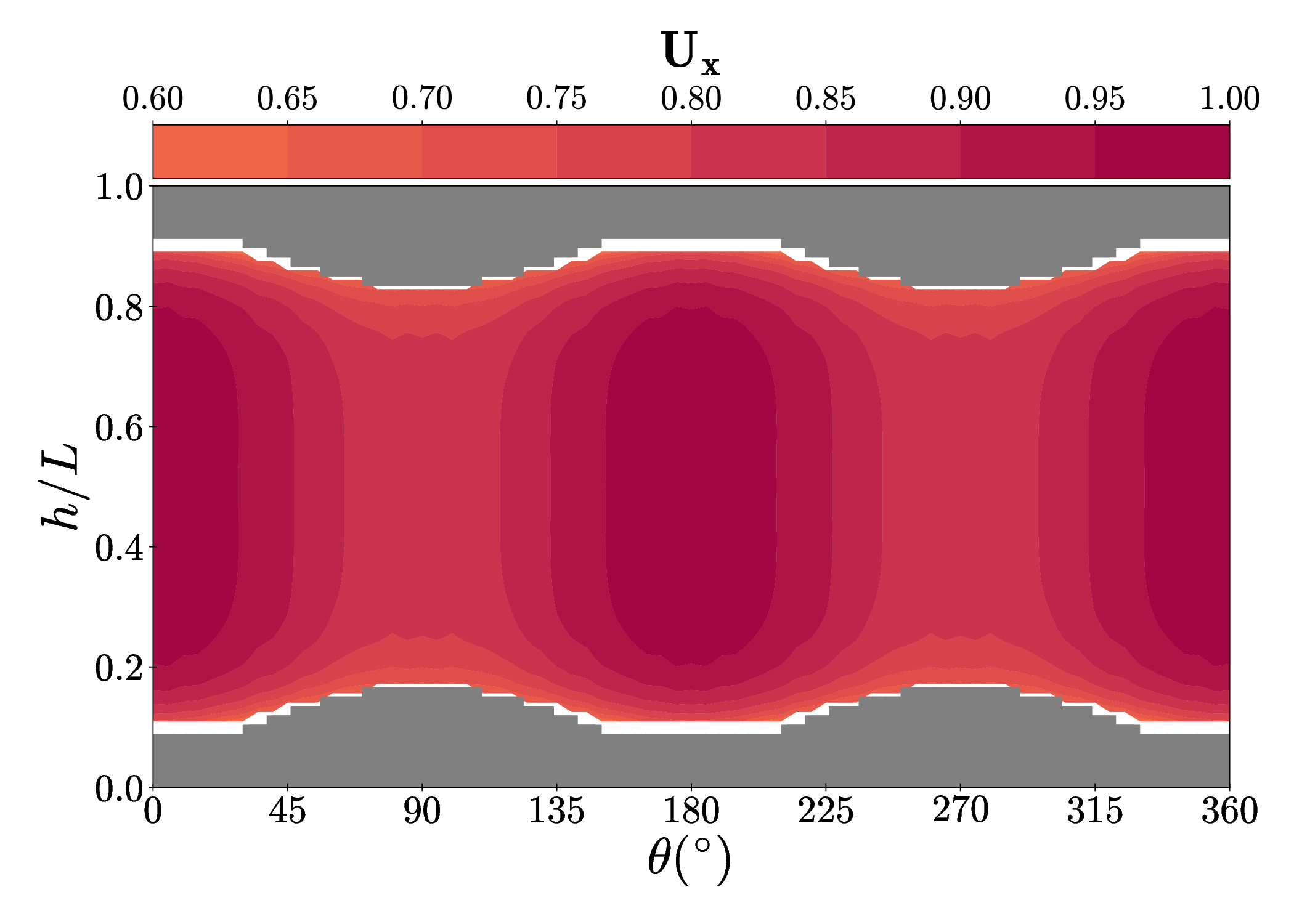}
         \caption{}
        \label{fig:VxPS}
    \end{subfigure}
    \hfill
     \centering
     \begin{subfigure}[b]{0.33\textwidth}
         \centering
         \includegraphics[width=\linewidth]{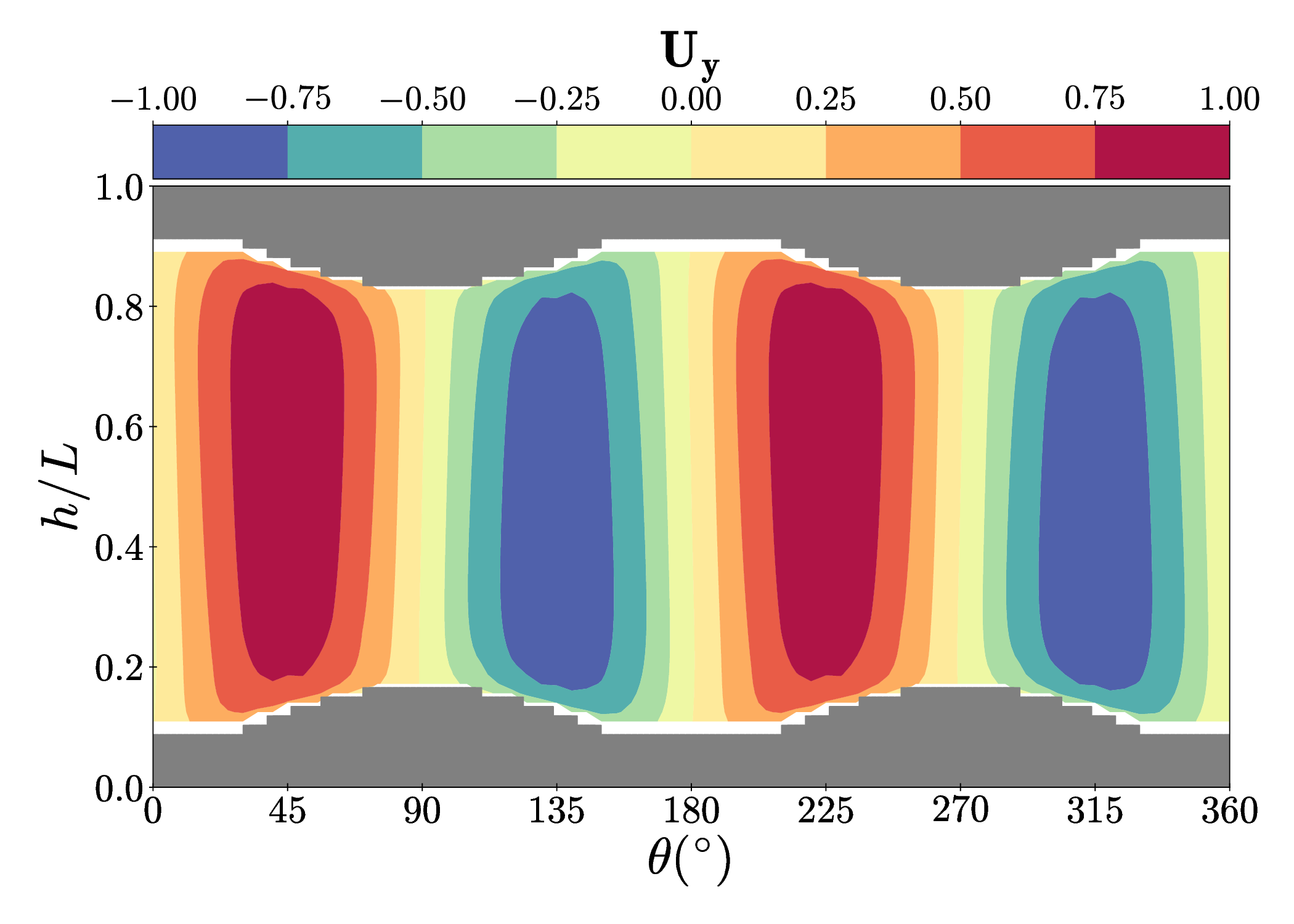}
         \caption{}
        \label{fig:VyPS}
    \end{subfigure}
    \hfill
    \centering
     \begin{subfigure}[b]{0.33\textwidth}
         \centering
         \includegraphics[width=\linewidth]{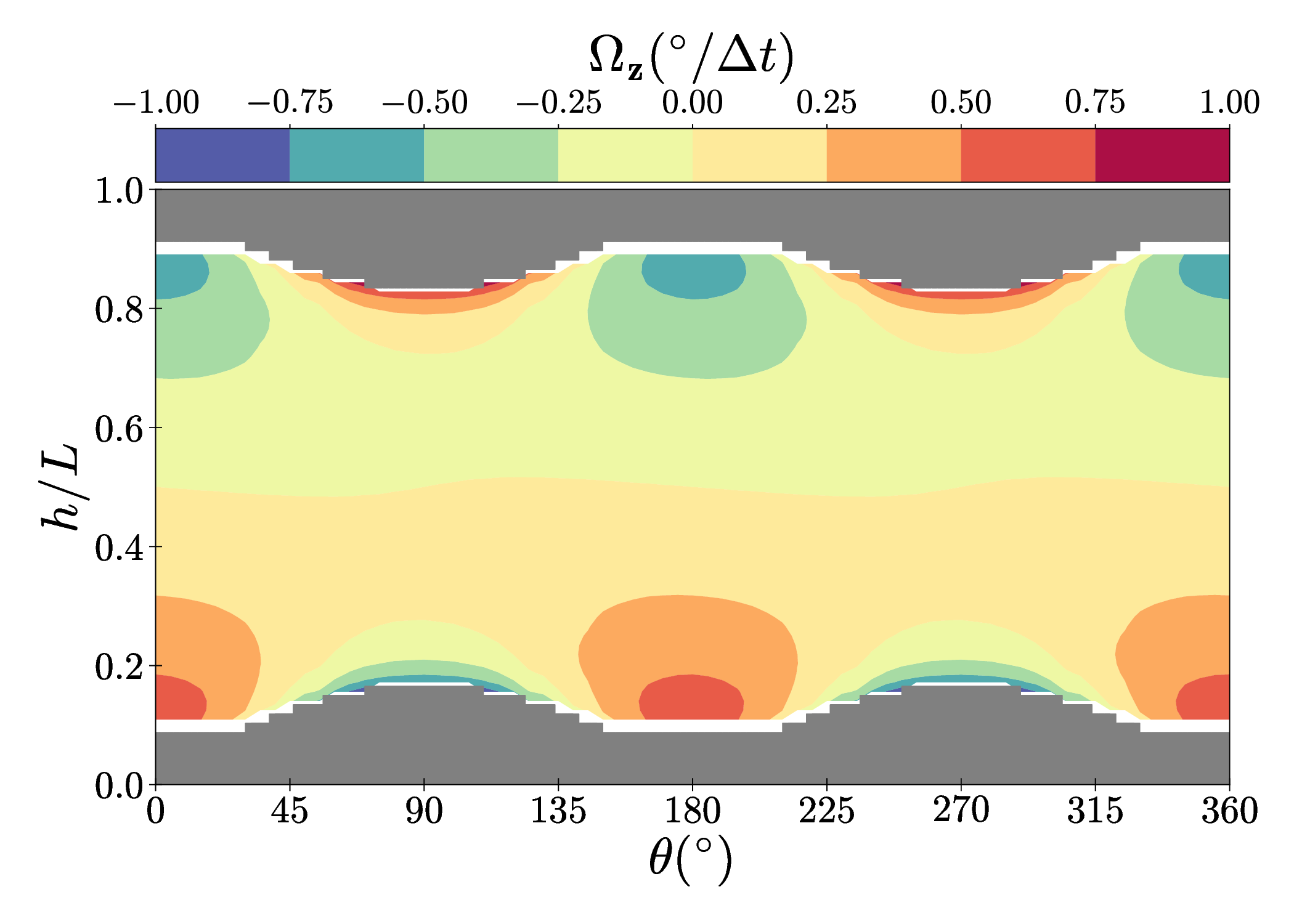}
         \caption{}
        \label{fig:WzPS}
    \end{subfigure}
    \hfill
     \begin{subfigure}[b]{0.33\textwidth}
         \centering
        \includegraphics[width=\textwidth]{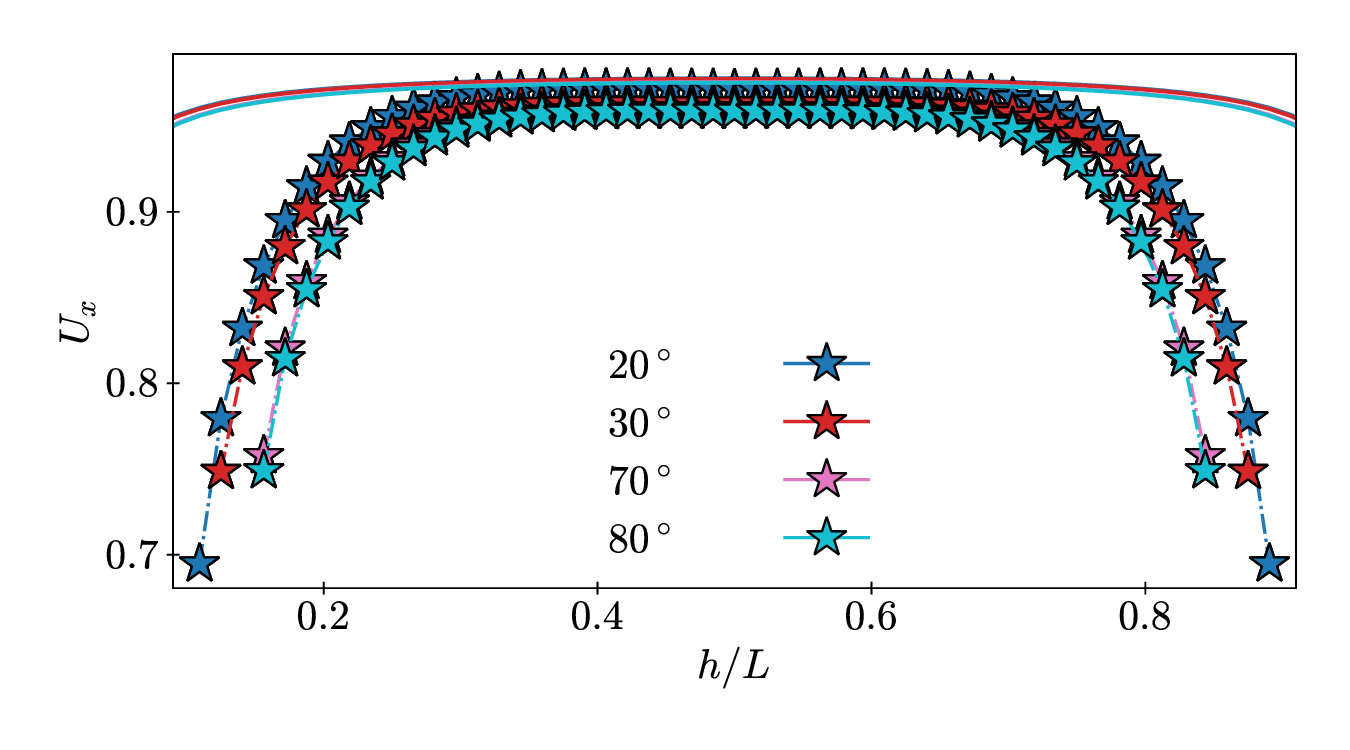}
        \caption{}
        \label{Ux_MS}
    \end{subfigure}
    \hfill
    \begin{subfigure}[b]{0.33\textwidth}
        \centering
        \includegraphics[width=\textwidth]{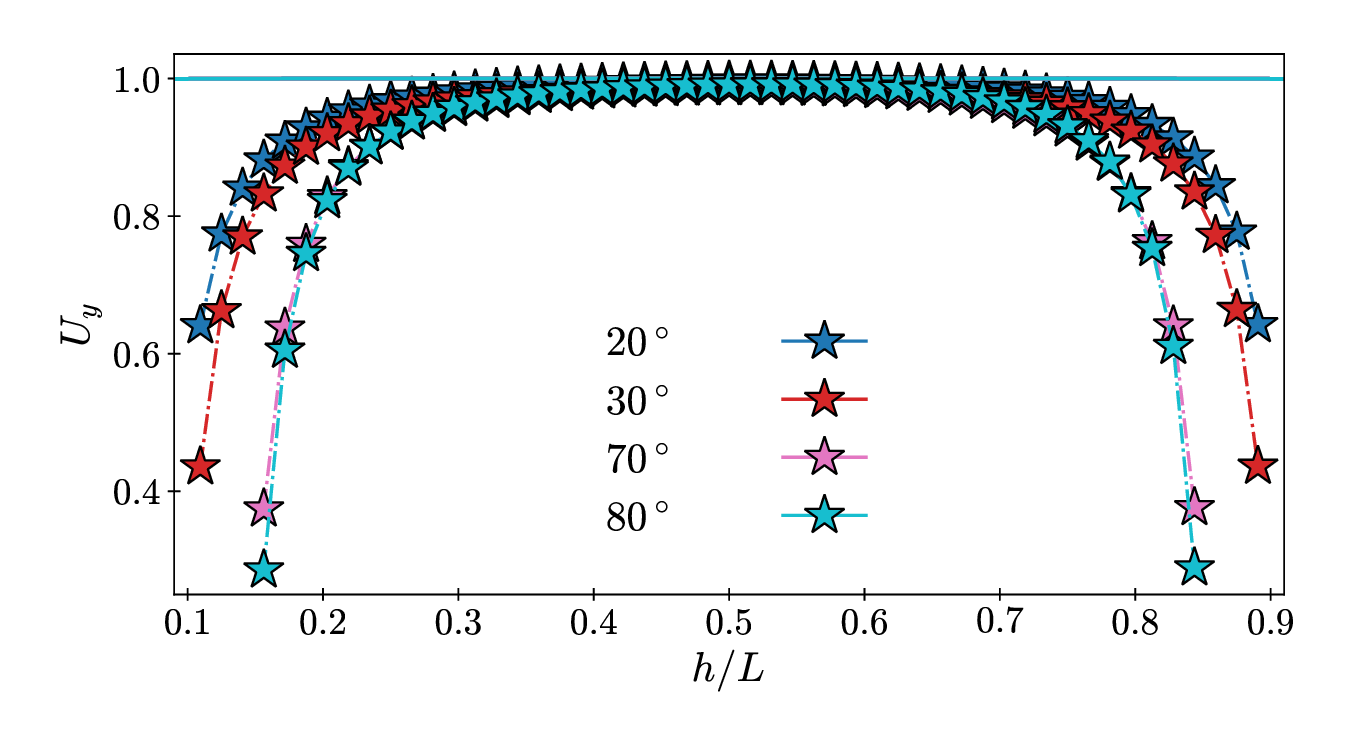}
        \caption{}
        \label{Uy_MS}
    \end{subfigure}
    \hfill
    \begin{subfigure}[b]{0.33\textwidth}
        \centering
        \includegraphics[width=\textwidth]{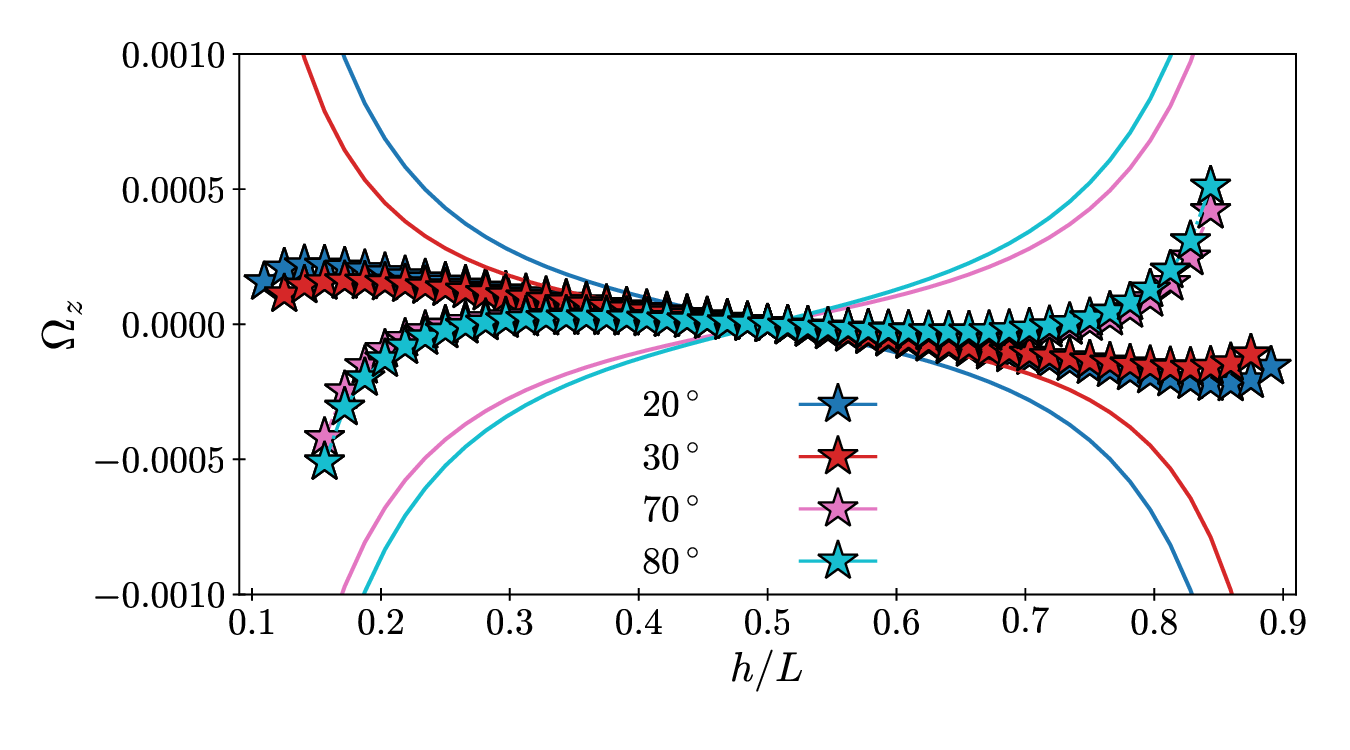}
         \caption{}
        \label{Wz_MS}
    \end{subfigure}
     \caption{\rev{Contour plots of (a) translational velocity $U_x$, (b) transverse velocity $U_y$ and (c) angular velocity $\varOmega_z$ in the $h-\theta$ space for a channel confined spheroid of aspect ratio $b/a\approx0.51$ at a confinement ratio CR$\approx0.15$. \rev{The colour bars are shown at the top of each plot. $h$ is normalised by the channel width $L$ and  $\theta$ is given in degrees ($ ^\circ$).} $\varOmega_z > 0$ corresponds to anti-clockwise rotation. In each figure, the inaccessible areas due to the finite size of the spheroid is coloured in grey. (d)-(f) Same data as line plots ($\bigstar$ symbols) along with results from the far field analysis (continuous lines) as a function of $h$ at various $\theta$. Results in (d) and (e) have been normalised with the translational velocity of the spheroid with same orientation ($\theta$) in an unbounded fluid. Plots (a) - (c) correspond to a ${\mathcal Re}\approx0.25$ where as plots (d) - (f) correspond to a ${\mathcal Re}\approx0.01$.}}
     \label{fig:eccentric}
\end{figure*}

\subsection{\textbf{Confined spheroid: eccentric positions in the channel}}
\label{sec:}

In this section, we relax the assumption of symmetric positioning and orientation of the spheroid and examine its dynamics for all possible positions and orientations within the channel. We define $h$ as the distance from wall 1 of the channel and $\theta$ as the orientation angle of the spheroid, as illustrated in Fig.~\ref{fig:prlschm}. To simplify the analysis, we restrict our investigation to the $xy$-plane at $z = L/2$. The translational velocity $U_x$ generally depends on both $h$ and $\theta$. In addition, the spheroid acquires a transverse velocity $U_y$ and an angular velocity $\varOmega_z$. The computed values of $U_x$, $U_y$ and $\varOmega_z$ as function of $h$ and $\theta$ are presented as contour plots in the $h-\theta$ plane in Fig.~\ref{fig:eccentric}. The simulation data along with theoretical predictions for $U_x, U_y$ and $\varOmega_z$ obtained from the far-field analysis are also shown as line plots in second row of  Fig.~\ref{fig:eccentric}.

Fig.~\ref{fig:VxPS} shows the translational velocity $U_x$ of the spheroid along the direction of the applied force (along the $x$-axis). The translational velocity is maximal when $h/L=0.5$ and $\theta = 0^\circ, 180^\circ$, corresponding to the end-on orientation of the spheroid positioned along the channel centreline. At other positions and orientations within the channel the translational velocity decreases because (1) the pressure drag increases as the projected surface area grows with $\theta$, and (2) the viscous drag increases due to hydrodynamic interactions with both the wall and the spheroid. The far-field predictions in Fig.~\ref{Ux_MS} show qualitative agreement with the full numerical results; however, the quantitative agreement deteriorates near the channel walls.

Fig.~\ref{fig:VyPS} illustrates the variation in the transverse velocity $U_y$, i.e., the velocity of the spheroid along the $y$-axis. The transverse velocity relative to the reference wall can be either positive or negative, and it changes sign every $90^\circ$ rotation of the spheroid, consistent with the symmetry of the system. The orientations at which the transverse velocity vanishes correspond to $\theta = 0^\circ, 90^\circ, 180^\circ,$ and $270^\circ$. At these angles, the external force $\bm{\hat{F}}$ is either fully parallel or fully perpendicular to $\hat{\bm{e}}$, and therefore the colloid does not migrate across the channel. In other words, hydrodynamic interactions with the wall in these configurations do not generate any transverse motion, in accordance with the reversibility constraints of Stokes flow.  As in the previous case for $U_x$, the comparison between the full numerical results and the far-field predictions (Fig.~\ref{Uy_MS}) is qualitative, with the agreement deteriorating near the walls. 

The angular velocity of the spheroid, $\varOmega_z$ in the $h-\theta$ phase space is shown in Fig.~\ref{fig:WzPS}. The rotation arises solely from the hydrodynamic torque on the particle, generated by the interplay between form drag, viscous drag, and the anisotropic shape of the spheroid. Depending on $h$ and $\theta$, the spheroid rotates clockwise ($\varOmega_z < 0$) or counterclockwise ($\varOmega_z > 0$) with the sign changing at the channel centreline. More importantly, the angular velocity also changes sign near the walls at $\theta = 45^\circ, 135^\circ$, etc., indicating that the direction of rotation as the spheroid approaches a wall depends strongly on its incident orientation. This orientation-dependent reversal of rotation leads to the glancing–reversing trajectories of spheroids discussed in the next section. The far-field predictions of the angular velocity shown in Fig.~\ref{Wz_MS} matches numerical simulations qualitatively, the asymmetry in the data in the numerical simulations is a finite Reynolds number effect (see section~\ref{sec:Re}).

\begin{figure*}
     \centering
     \begin{subfigure}[b]{0.4\textwidth}
         \centering
        \includegraphics[width=\textwidth]{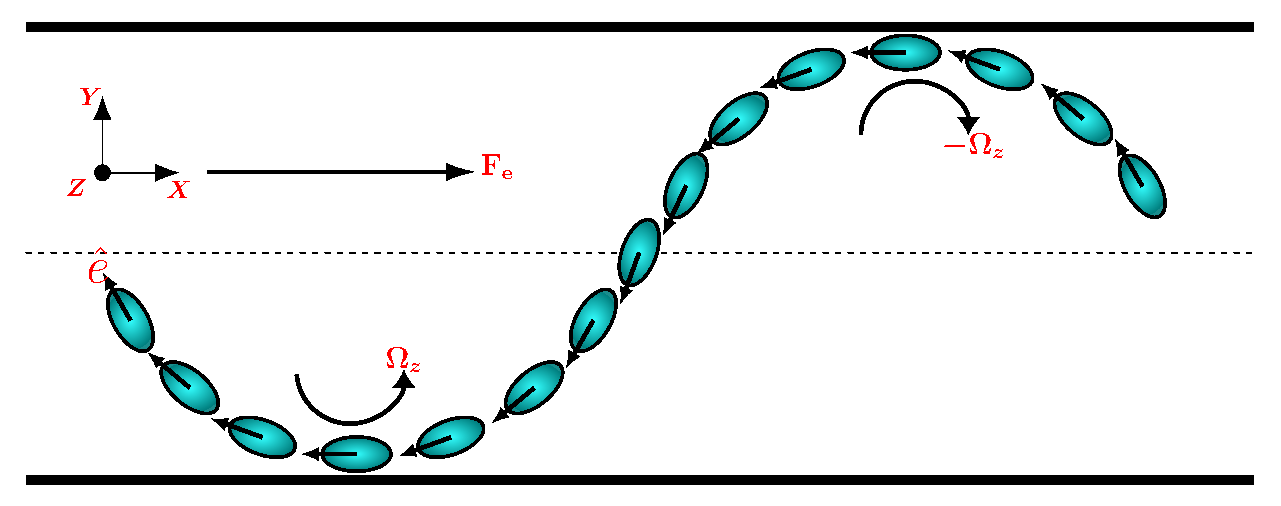}
        \caption{}
        \label{glsc}
    \end{subfigure}
    \hfill
    \centering
        \begin{subfigure}[b]{0.4\textwidth}
        \includegraphics[width=\textwidth]{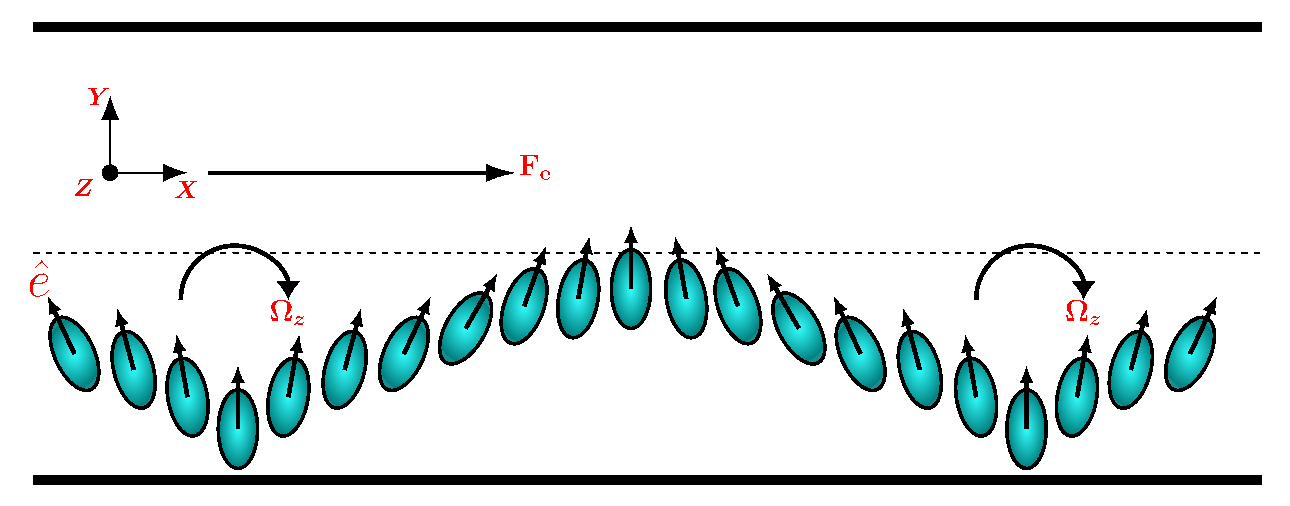}
        \caption{}
        \label{revsc}
    \end{subfigure}
    \hfill
         \centering
     \begin{subfigure}[b]{0.45\textwidth}
         \centering
         \includegraphics[trim = 0 0 40 0, clip, width=\linewidth]{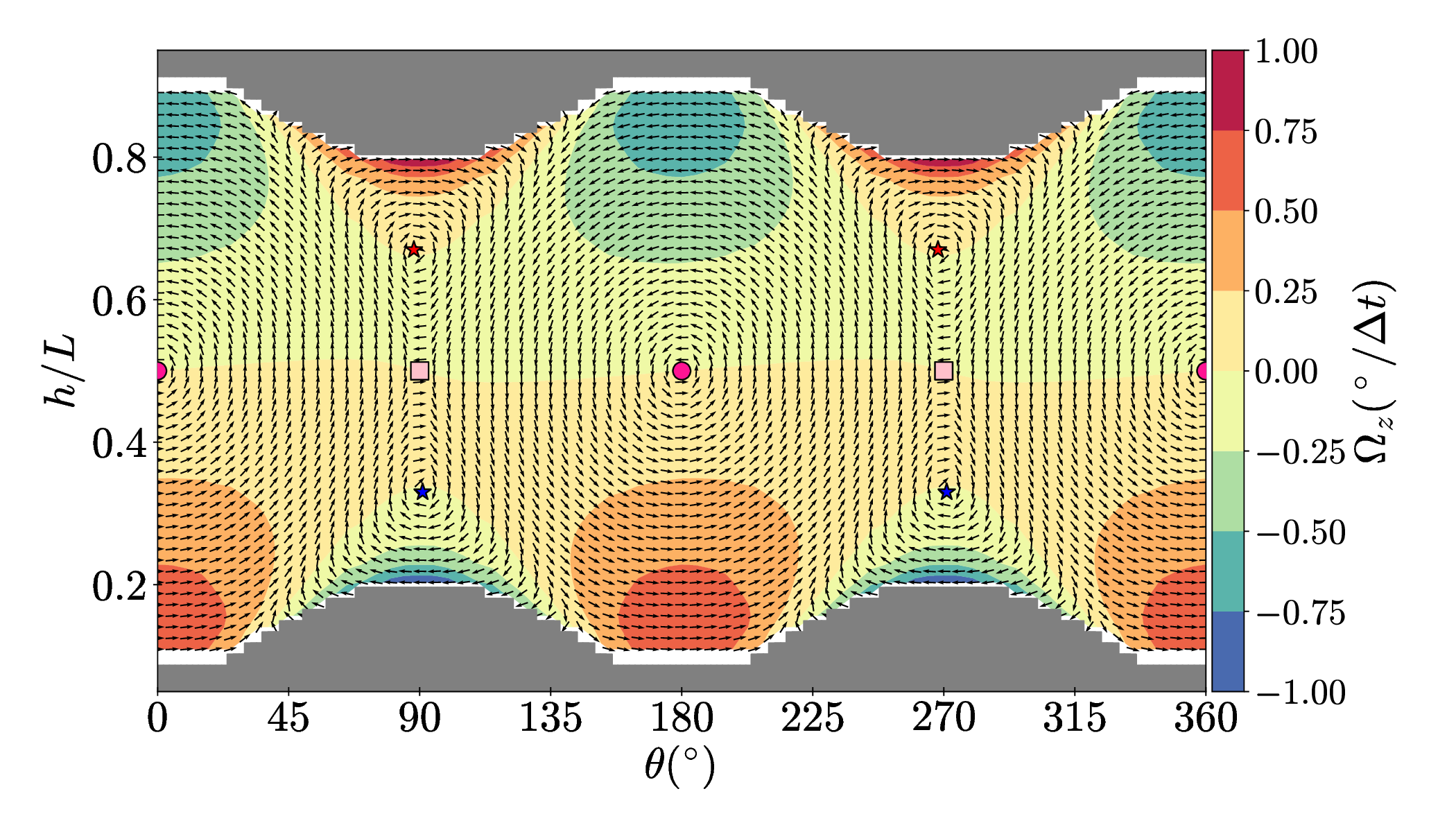}
         \caption{}
        \label{fig:036AR}
    \end{subfigure}
    \hfill
    \centering
     \begin{subfigure}[b]{0.45\textwidth}
         \centering
         \includegraphics[trim = 0 0 40 0, clip, width=\linewidth]{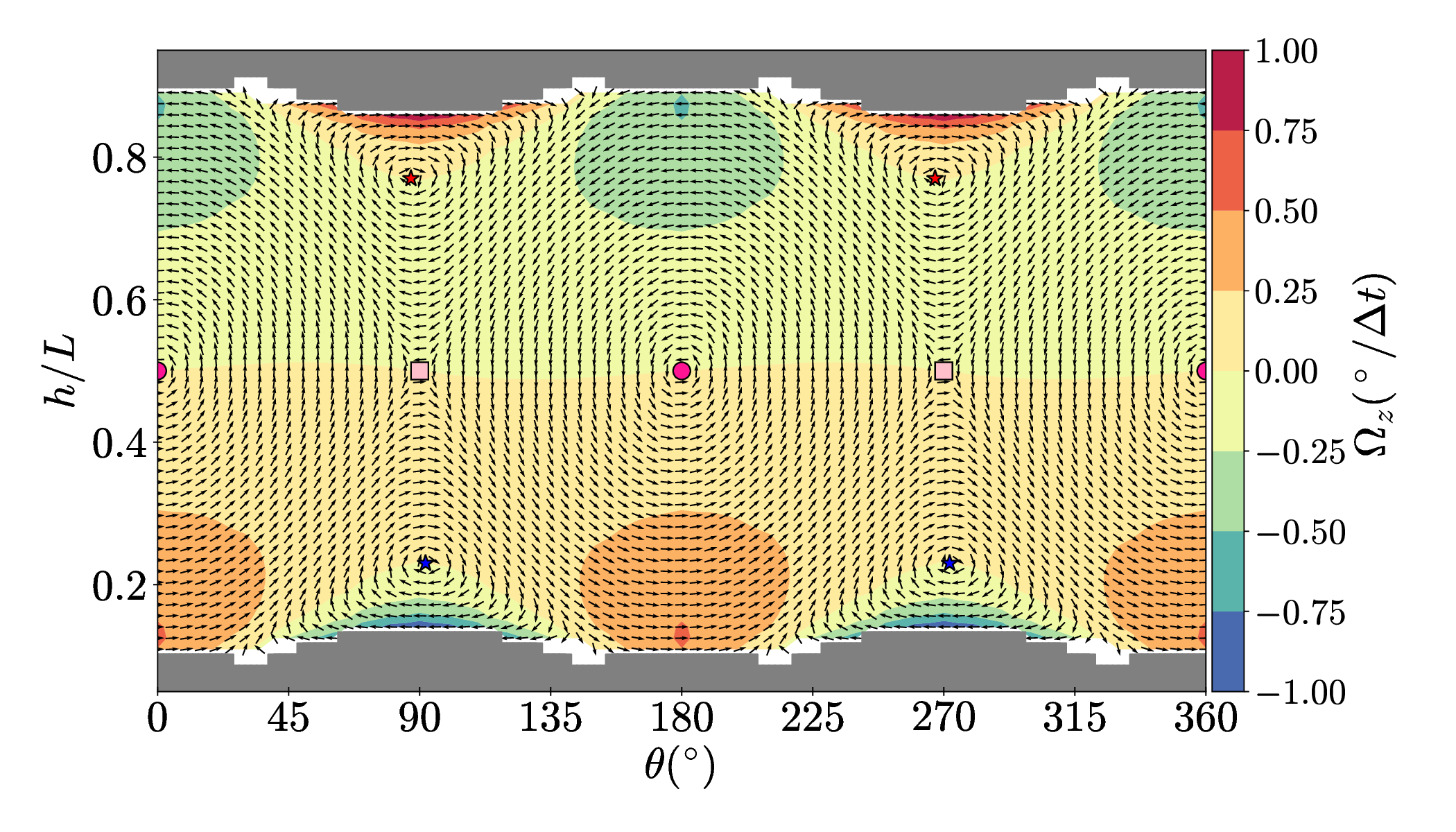}
         \caption{}
        \label{fig:07AR}
    \end{subfigure}
     \caption{Schematic representation of (a) glancing and (b) reversing spheroid. The dotted line represents the channel centreline. The instantaneous orientation, indicated by the arrow $\hat{\bm{e}}$, is provided as a visual guide to illustrate the angular rotation of the spheroid. Glancing (reversing) trajectories span the full (half) channel width, with the spheroid approaching each wall almost parallel (perpendicular) to it. \rev{(c)-(d) Velocity vectors in the $h-\theta$ phase space for a spheroid of aspect ratio $b/a\approx 0.36$ and $b/a\approx 0.7$. \rev{$h$ is normalised by the channel width $L$ and  $\theta$ is given in degrees ($ ^\circ$). The background is coloured with the angular velocity; $\varOmega_z > 0$ corresponds to anti-clockwise rotation. Black vector arrows indicate the  direction of the velocity of the spheroid in the phase space.} The fixed points are highlighted with different markers (see text). In each figure, the inaccessible areas due to the finite size of the spheroid is coloured in grey.}}
\end{figure*}

\subsection{\textbf{Glancing-reversing trajectories of the confined spheroid}}
\label{sec:}
We now analyse the trajectories of the confined spheroid as it translates through the channel under the action of the external force. Two types of trajectories are observed — glancing and reversing — similar to those reported for motion near a single wall~\cite{mitchell2015sedimentation}. These trajectories are illustrated schematically in Fig.~\ref{glsc} and Fig.~\ref{revsc}.

In a glancing trajectory (Fig.~\ref{glsc}), the spheroid grazes the channel wall with its long axis oriented nearly parallel to the wall. Due to the repulsive hydrodynamic interaction (i.e., the transverse velocity induced by the wall), the spheroid begins to move away from the wall while undergoing anticlockwise rotation. This drift causes the spheroid to approach the opposite wall, where it is again repelled while rotating clockwise, eventually returning to the first wall. Thus, over one full cycle, the spheroid spans the entire channel width, with its orientation oscillating between angles less than and greater than $180^\circ$. In contrast, during a reversing trajectory (Fig.~\ref{revsc}), the spheroid remains near one wall and never crosses the channel centreline. In this case, the spheroid approaches the wall with its long axis nearly perpendicular to it, and hydrodynamic interactions cause its orientation to oscillate between angles less than and greater than $90^\circ$ during each cycle. Thus, although both types of trajectories are oscillatory, the resulting motions are qualitatively different. 

The origin of these two types of channel-confined spheroid trajectories can be understood by examining the corresponding phase-space dynamics. The velocity vectors in the $h-\theta$ phase space for spheroids of two different aspect ratios are shown in Fig.~\ref{fig:036AR} and Fig.~\ref{fig:07AR}. 
The phase space is coloured by the angular velocity $\varOmega_z$. As expected for a low–Reynolds-number system, the phase-space trajectories form closed loops, consistent with the oscillatory motion of the spheroid in the channel. With these phase portraits, we can identify the origin of the two distinct classes of trajectories.

The first type of trajectory are closed loops centred about $h/L = 0.5$ and $\theta=180^{\circ}$ (marked $\CIRCLE$ in the figures); this set corresponds to the glancing trajectories.  Glancing trajectories of opposite sense of rotation are centred at $0^\circ$ (or $360^\circ$ equivalently). These trajectories occupy the entire width of the channel, thus crossing the channel centre line where the angular velocity changes sign. Glancing occurs when a spheroid approaching the bottom (or top) wall with its long axis nearly parallel to the wall undergoes anticlockwise (clockwise) rotation and subsequently drifts toward the opposite wall. There exist infinitely many glancing trajectories, each individual trajectory is uniquely determined by the initial position and orientation of the spheroid. 

The second type, the reversing trajectories are sandwiched between the glancing trajectories. They are closed loops, centred around values of $h/L$ close to either the top or bottom wall at $\theta=90^{\circ}$ (marked $\star$ in the figures). They occupy at most one half of the channel width. A pair of reversing trajectories with opposite sense of rotation exists symmetrically about the channel centreline (on either side of the symbol $\blacksquare$). An equivalent pair is centred about $\theta=270^\circ$.



In general, the reversing trajectories are more skewed in phase space compared to the glancing trajectories. Importantly, these qualitative features are independent of the spheroid aspect ratio (see Fig.~\ref{fig:036AR} and Fig.~\ref{fig:07AR}). Thus, glancing and reversing trajectories are intrinsic characteristics of channel-confined, force-driven spheroidal particles.

The end-on configuration of a prolate spheroid translating along the centreline of the channel ($h/L = 0.5, \theta = 0^\circ$), discussed in Sec.~\ref{sec:centre_confined}, forms the centre of the glancing trajectories. This configuration corresponds to a neutrally stable point in the phase space. In addition to this, the phase space contains a pair of additional, neutrally stable centres — these are the centre points of the reversing trajectories — located on either side of the channel centreline and closer to the top and bottom walls. On the other hand, the broadside-on configuration at the channel centreline ($h/L = 0.5, \theta = 90^\circ$) is a saddle point formed between the pair of glancing trajectories (of opposite sense of rotation) and the pair of reversing trajectories (also of opposite sense of rotation), marked $\blacksquare$ in the figures.~Consequently, the broadside-on state of a channel-confined spheroid at the centreline, is an unstable configuration; any small perturbation in position or orientation drives the spheroid away from this state and into either a glancing or a reversing trajectory.



\begin{figure*}
     \centering
     \begin{subfigure}[b]{0.338\textwidth}
         \centering
         \includegraphics[width=\linewidth]{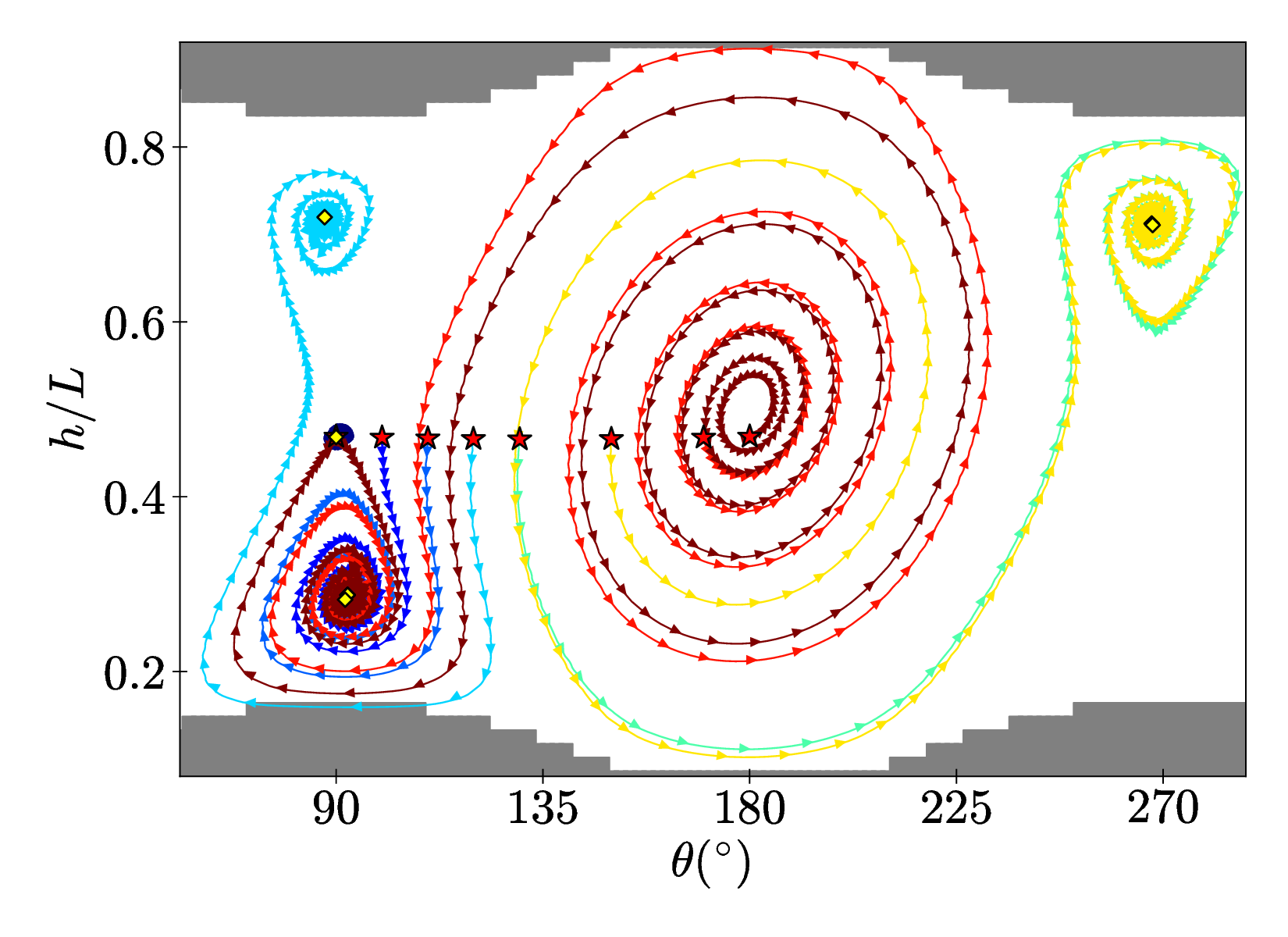}
         \caption{}
        \label{fig:150re}
    \end{subfigure}
    \hfill
     \centering
     \begin{subfigure}[b]{0.32\textwidth}
         \centering
         \includegraphics[trim = 30 20 250 0, clip,width=\linewidth]{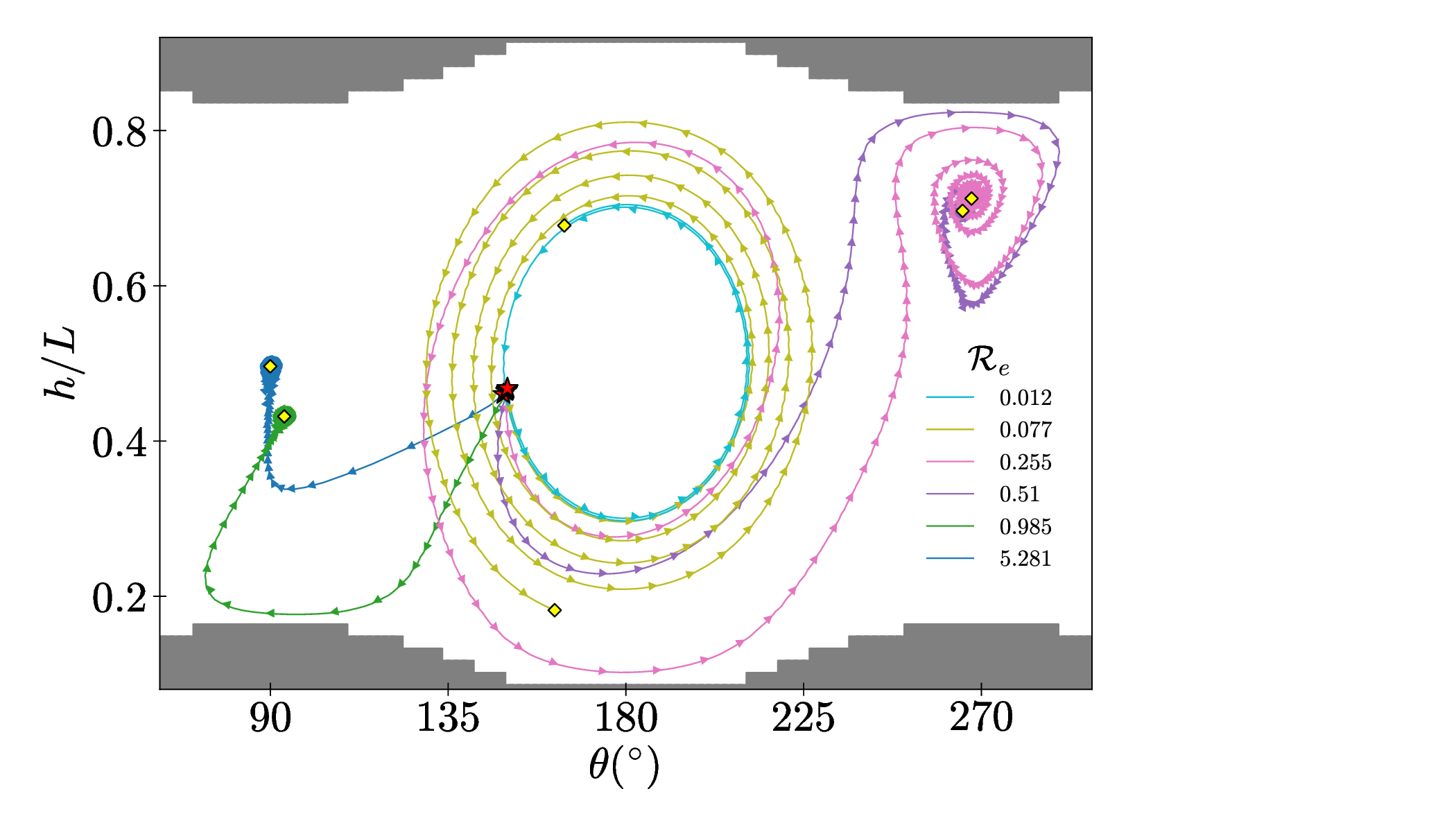}
         \caption{}
        \label{fig:170re}
    \end{subfigure}
    \hfill
    \centering
     \begin{subfigure}[b]{0.32\textwidth}
         \centering
         \includegraphics[trim = 30 20 250 0, clip,width=\linewidth]{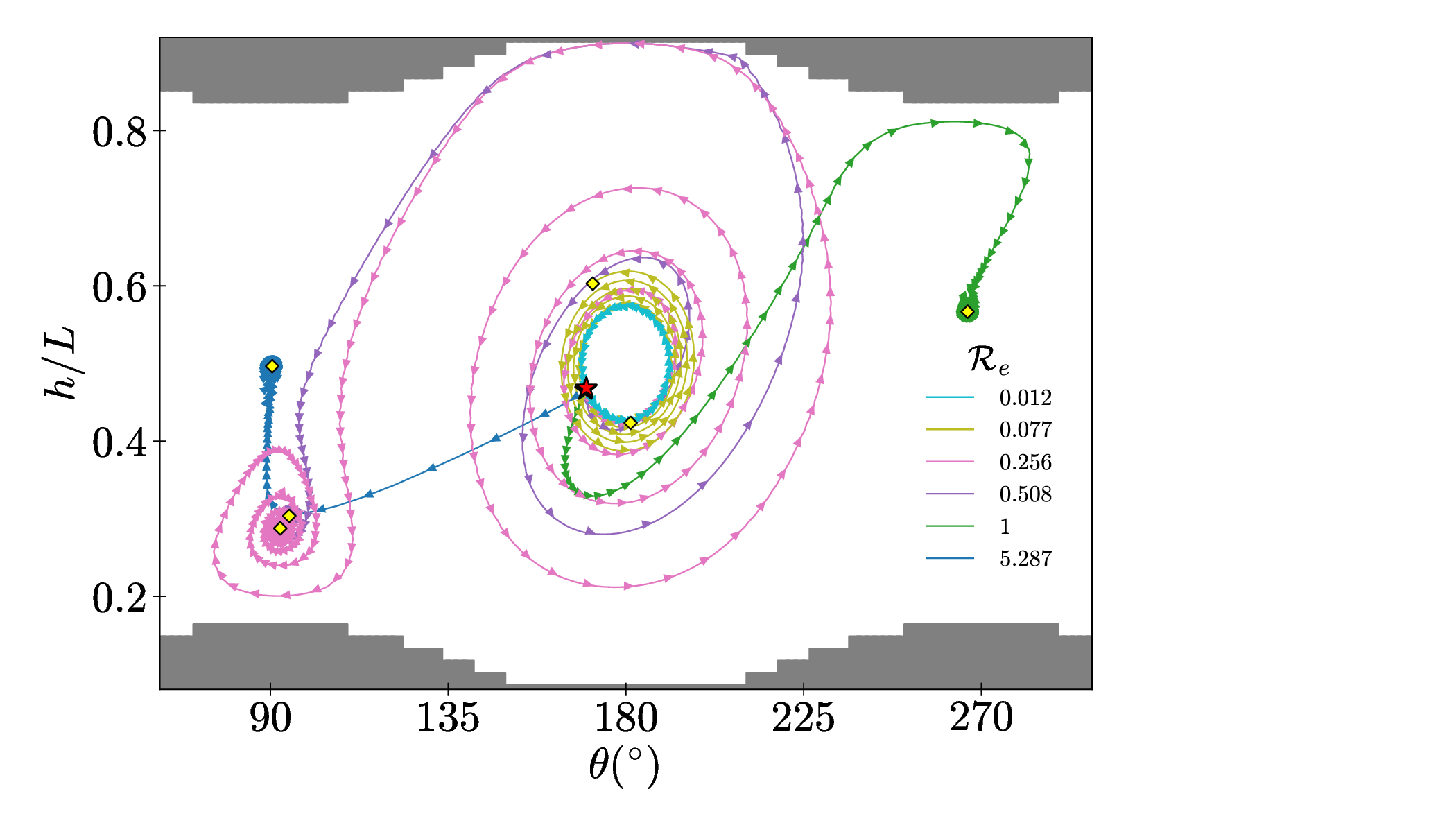}
         \caption{}
        \label{fig:0225theta}
    \end{subfigure}
     \caption{(a) Phase space trajectories of the spheroid  at ${\mathcal Re}\approx0.2$. The trajectories are obtained by initialising the spheroid at $h/L\approx0.47$ and $\theta$ varied in $90^\circ  - 180^\circ$. Phase space trajectories of the spheroid  at various ${\mathcal Re}$ (b) for initial orientation $\theta = 150^\circ$ and (c) initial orientation $\theta = 170^\circ$ for initial $h/L \approx0.47$. In each figure, the initial and final points are highlighted with $\bigstar$ and $\diamond$ respectively. The final point corresponds to the last data point obtained in the simulation lasting $6\times 10^6$ iteration steps, and may not necessarily correspond to a fixed point.}
\end{figure*}

\subsection{\textbf{Effect of fluid inertia and resulting bifurcations}}
\label{sec:Re}

In the results discussed so far, we have neglected the effects of fluid inertia and assumed the Stokes regime. We now relax this assumption and examine the influence of  fluid inertia on the dynamics of a channel-confined spheroid. We fix the aspect ratio of the prolate spheroid at  $b/a\approx0.51$ and vary the Reynolds number ${\mathcal Re}$. \rev{The reported ${\mathcal Re}$ is based on the steady state velocity of the translating spheroid.} Note that the quasi-static simulations cannot be used to study the effect of inertia; we obtain the complete trajectory of the spheroid using LBM simulations. 


The phase-space trajectories in the $h-\theta$ plane for a channel-confined spheroid at ${\mathcal Re}\approx0.2$ are shown in Fig.~\ref{fig:150re}. These trajectories are generated by initialising the spheroid at $h\approx0.47$, close to the channel centreline, and varying the initial orientations $\theta$  in the range $90^\circ - 180^\circ$. The initial and final points are marked with $\star$ and  $\diamond$ symbols respectively. Unlike the dynamics in the Stokes regime, the paths that the spheroid takes in phase-space are no longer closed loops. The glancing trajectories spiral outward and open up, and as a result, each glancing trajectory ultimately merges smoothly into a reversing trajectory. The reversing trajectories, in turn, spiral inward toward their centre point. Consequently, a spheroid that initially follows a glancing path will eventually transition into a reversing trajectory and then relax toward a broadside-on configuration ($\theta=90^\circ$) near either the top or bottom wall.

These qualitative changes arise from the introduction of fluid inertia. At finite ${\mathcal Re}$, the flow no longer adjusts instantaneously to the particle configuration, and inertial forces develop that break the neutrally stable closed loops characteristic of Stokes flow. Glancing trajectories experience a net outward drift away from the centreline, while reversing trajectories experience an inward drift toward orientations near $\theta=90^\circ$. As a result, the end-on configuration at the centre point, $h/L = 0.5, \theta=180^{\circ}$ loses neutral stability, and the system develops four stable broadside-on fixed points — two near the upper wall and two near the lower wall. The saddle node at  $h/L = 0.5, \theta=90^{\circ}$ remains an unstable fixed point at this Reynolds number (${\mathcal Re} \approx 0.2$). However, the phase-space structure undergoes significant changes as ${\mathcal Re}$ is increased further.


To systematically examine the influence of fluid inertia, we consider two initial orientations, $\theta = 150^\circ$ and $\theta=170^\circ$ at $h/L\approx0.47$, and vary the Reynolds number ${\mathcal Re}$. The resulting trajectories in the $h-\theta$ phase space are shown in Fig.~\ref{fig:170re} and~\ref{fig:0225theta} respectively. The dynamics exhibit pronounced nonlinearity: the outward spiralling characteristic of glancing trajectories and the inward spiralling associated with reversing trajectories are visible only at relatively small ${\mathcal Re}$. As ${\mathcal Re}$ increases beyond $\mathcal{O}(1)$, additional fixed points emerge, and the trajectories bear little resemblance to the aforementioned glancing and reversing paths observed in the Stokes regime. At the highest Reynolds number considered in this study (${\mathcal R}e\approx 5$), the spheroid rapidly settles into a stable broadside-on configuration at the channel centreline ($h/L = 0.5, \theta=90^\circ$), similar to that of an unconfined spheroid \cite{khayat1989inertia,huang1998direct,pan2002direct,feng1994direct,dabade2015effects}. Thus, even modest levels of inertia can fundamentally alter the topology of the phase space and promote stable broadside-on alignment. 

\begin{figure}
     \centering
         \includegraphics[width=1.02\linewidth]{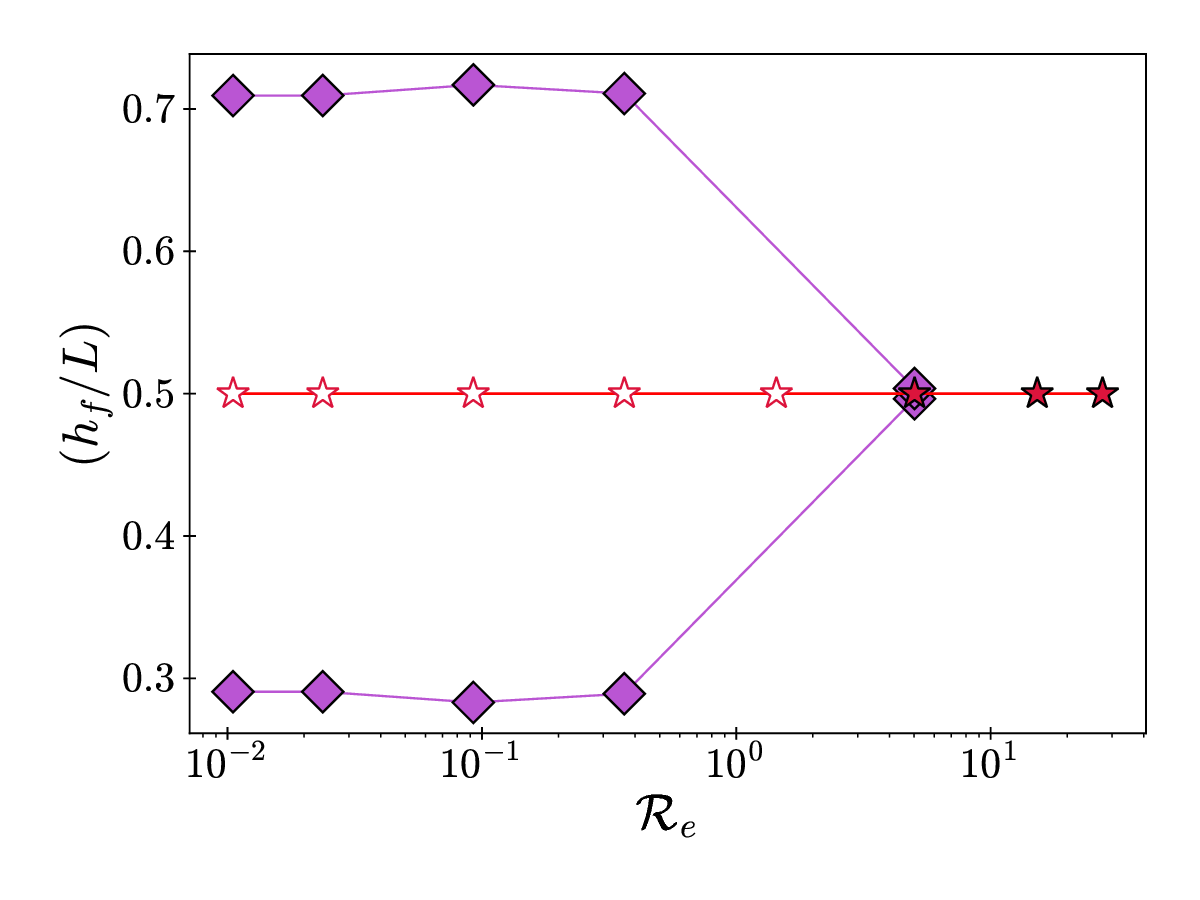}
         \caption{The bifurcation diagram shows effect of ${\mathcal Re}$ on the final position along the channel width $(h_f/L)$. The stable (unstable) points are indicated by filled (open) symbols. }
        \label{fig:bifurcation}
\end{figure}

To map the dependence of the stable fixed points on the Reynolds number, we construct the bifurcation diagram shown in Fig.~\ref{fig:bifurcation}. The plot corresponds to a prolate spheroid of aspect ratio $b/a\approx0.51$, and in this diagram the final equilibrium position, denoted by $(h_f/L)$ is plotted as a function of the Reynolds number ${\mathcal Re}$. The diagram is obtained by initialising the spheroid at various positions $(h/L)$ across the channel width and allowing it to evolve to its steady state. For ${\mathcal Re} > \mathcal{O}(1)$ a single stable solution exists at $h_f/L = 0.5$, corresponding to broadside-on translation at the channel centreline. In contrast, for small Reynolds numbers ${\mathcal Re} < \mathcal{O}(1)$, this centreline state is unstable, and stable broadside-on motion occurs near the channel walls. This transition indicates the presence of a bifurcation at ${\mathcal Re} \approx \mathcal{O}(1)$. A detailed characterisation of the mathematical nature of this bifurcation requires further analytical and numerical investigation and is beyond the scope of the present work.

\rev{The presence of spiralling trajectories in the phase space indicates that fluid inertia also breaks the symmetry in the translational and angular kinematics of the spheroid with respect to the channel centreline. The extent of asymmetry is proportional to the Reynolds number ${\mathcal Re}$. We find that both components of translational velocity and the angular velocity, $U_x$, $U_y$ and $\Omega_z$ are visibly not symmetric with respect to the channel centreline when quasi-static simulations are performed at ${\mathcal Re}\approx \mathcal{O}(0.1)$.  With regard to the final configuration of the particle, previous studies have shown that fluid-inertia-induced torque on non-spherical particles cannot be neglected even at Reynolds numbers much smaller than unity \cite{dabade2015effects,sheikh2020importance}. The preferred orientation of a translating  spheroid due to finite inertia is at~$\theta = 90^\circ$, \textit{i.e.,} the broadside-one configuration \cite{khayat1989inertia, huang1998direct, pan2002direct, feng1994direct, dabade2015effects, jiang2024prolate, ardekani2016numerical}. Our results show that, even in confinement this result holds; but the location of the equilibrium positions can be either at the channel centre line or near the walls depending upon ${\mathcal Re}$ as shown in Fig.~7. Hence, the approximate results obtained by~\citet{mitchell2015sedimentation} which account for the wall induced hydrodynamic interaction can be substantially altered due to the presence of a small but finite Reynolds number.}

\section{Conclusions}
\label{sec:conclusions}

Anisotropic particles appear widely in soft-matter systems and complex fluids, and even in simple Newtonian fluids they exhibit remarkably rich dynamics. In this work, we used lattice Boltzmann simulations to examine the motion of a force-driven spheroid and complemented these simulations with an analytical approach based on the superposition principle and far-field hydrodynamics. By systematically varying the particle shape, degree of confinement, and fluid inertia, we demonstrated how each of these factors influences the particle dynamics.

In unconfined conditions, we found that the shape corresponding to the maximum translational velocity for a fixed particle volume is not a sphere. Instead, the highest velocity occurs for a prolate spheroid of aspect ratio $b/a \approx 0.512$ in an end-on configuration and for an oblate spheroid of aspect ratio $b/a \approx 1.514$ in a broadside-on configuration. A similar trend persists in confined geometries, where the optimal shape typically corresponds to oblate spheroids. When a spheroid is placed at an off-centre position or orientation within the channel, transverse and angular velocities are generated, and the coupling between translation and rotation produces oscillatory trajectories. Two distinct types of oscillatory dynamics — glancing and reversing — were identified. In a glancing trajectory, the spheroid oscillates between the two channel walls with its long axis nearly parallel to the wall near each encounter. In contrast, reversing trajectories are confined to one half of the channel width, with the spheroid approaching the wall in a broadside-on configuration.

These oscillatory motions appear as initial-condition–dependent closed loops in phase space under Stokes flow. However, the inclusion of fluid inertia breaks these closed loops. At small but finite Reynolds numbers, glancing trajectories spiral outward and merge into reversing trajectories, which spiral inward toward a fixed point corresponding to a broadside-on orientation near the walls. As the Reynolds number increases further, the phase-space structure undergoes dramatic changes: new fixed points emerge, existing ones change stability, and the overall dynamical landscape becomes more intricate. We have characterised these transitions using a bifurcation diagram. While our study is primarily computational, it highlights the need for future investigations from a nonlinear dynamical perspective to fully unravel the complexity of this system.

In summary, our work reveals the rich and intricate dynamics associated with the classical problem of a force-driven spheroid in a confined channel. Further analytical and numerical studies are needed to fully understand the nonlinear behaviour uncovered here. Nevertheless, our systematic analysis offers predictions that can be readily tested in microfluidic experiments and provides guidance for optimising the shape of drug-delivery particles, illustrating the broad possibilities that nonspherical particles bring to applications in transport and design.

\section*{\rev{Conflicts of interest}}
\rev{There are no conflicts of interest to declare.}

\section*{Acknowledgements}
Computational facilities at IIT Madras are duly acknowledged. SPT acknowledges the project support by I-Hub Foundation for Cobotics (IHFC), IITD and the travel support by IoE, IIT Madras.
This work used the ARCHIE-WeSt High-Performance Computer (\href{https://www.archie-west.ac.uk}{www.archie-west.ac.uk}) based at the University of Strathclyde.
For the purpose of complying with UKRI's open access policy, the authors have applied a Creative Commons Attribution (CC BY) license to any Author Accepted Manuscript version arising from this submission.

\bibliography{biblio.bib}


\end{document}